\documentclass[letterpaper]{article} % DO NOT CHANGE THIS
\usepackage{aaai2026}  % DO NOT CHANGE THIS
\usepackage{times}  % DO NOT CHANGE THIS
\usepackage{helvet}  % DO NOT CHANGE THIS
\usepackage{courier}  % DO NOT CHANGE THIS
\usepackage[hyphens]{url}  % DO NOT CHANGE THIS
\usepackage{graphicx} % DO NOT CHANGE THIS
\usepackage{natbib}  % DO NOT CHANGE THIS AND DO NOT ADD ANY OPTIONS TO IT
\usepackage{caption} % DO NOT CHANGE THIS AND DO NOT ADD ANY OPTIONS TO IT
\usepackage{algorithm}
\usepackage{algorithmic}

\usepackage{newfloat}
\usepackage{listings}
\DeclareCaptionStyle{ruled}{labelfont=normalfont,labelsep=colon,strut=off} % DO NOT CHANGE THIS
\floatstyle{ruled}
\newfloat{listing}{tb}{lst}{}
\floatname{listing}{Listing}
\usepackage{booktabs}
\usepackage{multirow}
\usepackage{makecell}

\usepackage{amssymb}
\usepackage{amsmath}
\usepackage{graphicx}

\title{Beyond Content: A Comprehensive Speech Toxicity Dataset and Detection Framework Incorporating Paralinguistic Cues}
\author{
    Zhongjie Ba\textsuperscript{\rm 1,\rm 2},
    Liang Yi\textsuperscript{\rm 1},
    Peng Cheng\textsuperscript{\rm 1,\rm 2} \thanks{The corresponding author.},
    Qingcao Li\textsuperscript{\rm 3,\rm 1},
    Qinglong Wang\textsuperscript{\rm 1,\rm 2},
    Li Lu\textsuperscript{\rm 1,\rm 2}
}

\affiliations{
    \textsuperscript{\rm 1}The State Key Laboratory of Blockchain and Data Security, Zhejiang University\\
    \textsuperscript{\rm 2}Hangzhou High-Tech Zone (Binjiang) Institute of Blockchain and Data Security, Hangzhou, Zhejiang, China\\
    \textsuperscript{\rm 3}School of Cyber Science and Engineering, Nanjing University of Science and Technology\\
    \{zhongjieba,yiliang,peng\_cheng\}@zju.edu.cn,
    liqingcao@njust.edu.cn,
    \{qinglong.wang,li.lu\}@zju.edu.cn
}

\usepackage{bibentry}
\begin{document}

\maketitle

\begin{abstract}
Toxic speech detection has become a crucial challenge in maintaining safe online communication environments. However, existing approaches to toxic speech detection often neglect the contribution of paralinguistic cues, such as emotion, intonation, and speech rate, which are key to detecting speech toxicity. Moreover, current toxic speech datasets are predominantly text-based, limiting the development of models that can capture paralinguistic cues. To address these challenges, we present ToxiAlert-Bench, a large-scale audio dataset comprising over 30,000 audio clips annotated with seven major toxic categories and twenty fine-grained toxic labels. Uniquely, our dataset annotates toxicity sources—distinguishing between textual content and paralinguistic origins—for comprehensive toxic speech analysis. Furthermore, we propose a dual-head neural network with a multi-stage training strategy tailored for toxic speech detection. This architecture features two task-specific classification headers: one for identifying the source of sensitivity (textual or paralinguistic), and the other for categorizing the specific toxic type. The training process involves independent head training followed by joint fine-tuning to reduce task interference. To mitigate data class imbalance, we incorporate class-balanced sampling and weighted loss functions. Our experimental results show that leveraging paralinguistic features significantly improves detection performance. Our method consistently outperforms existing baselines across multiple evaluation metrics, with a 21.1\% relative improvement in Macro-F1 score and a 13.0\% relative gain in accuracy over the strongest baseline, highlighting its enhanced effectiveness and practical applicability.
\end{abstract}


\begin{links}
    \link{Code}{https://github.com/yiliang-la/ToxiAlert}
\end{links}

\section{Introduction}
Toxic speech, as part of toxic behaviors, can occur virtually and physically, resulting in a negative psychological impact~\cite{nada2023lightweight}. ``Toxic speech" often includes hostile intent that is threatening, abusive, discriminatory, etc.~\cite{garg2023handling, fortuna2020toxic}. Such behavior can target individuals or groups, leading to severe outcomes such as cyberbullying, harassment, and the spread of discriminatory ideas. 

Voice-based social platforms have surged in recent years, amplifying the risk of spreading toxicity through audio. Popular social platforms like Twitter and Facebook have mature text-based content moderation systems~\cite{zeng2024shieldgemma, inan2023llama}. However, voice-involving social platforms for live streaming, multiplayer online gaming, or voice/video chatting, such as Twitch and Slack, require more than text-based moderation~\cite{hamilton2014streaming}. Researchers have discovered that when it comes to detecting intents embedded within audio signals, purely text-based models are not sufficient~\cite{lin2022toxic}. The combination of verbal and non-verbal cues can express some toxic intentions. For example, explicit adult content can be conveyed through non-verbal cues such as moaning or Autonomous Sensory Meridian Response (ASMR)-like sounds, which can bypass text-based detection. Therefore, \emph{how to leverage paralinguistic cues for effective detection of toxic speech urges more study.}

Recent studies have realized the importance of acoustic features other than semantic content in toxic speech identification. Relevant studies can be divided into three categories: Generic Acoustic-Based~\cite{Yousefi2021Audio, ghosh22b_interspeech}, Feature Fusion-Based~\cite{lin2022toxic, rana2022emotion, mandal2024attentive}, and Textual Task-Assisted Multi-task Learning~\cite{nada2023lightweight, liu2024enhancing2,nandwana2024voice}. 

Despite these advances, several critical limitations hinder progress in paralinguistic-aware toxic speech detection. 

\noindent\textbf{Lack of Suitable Datasets. }The scarcity of publicly available datasets poses a significant barrier to research development. Among existing works, only DeToxy~\cite{ghosh22b_interspeech} has released a public dataset (i.e., DeToxy-B), but its toxicity classification is solely based on textual content. It lacks samples where toxicity originates from paralinguistic cues~\cite{scherer1973voice} alone or from both textual and paralinguistic sources combined. Consequently, existing public datasets are insufficient to support the development of detection systems capable of identifying paralinguistic-based toxicity. While some studies focus on non-textual information of audio signals as sources of toxicity and develop corresponding detection systems, their datasets remain private and undisclosed, with unclear and non-transparent construction methodologies. 

\noindent\textbf{Technical Limitations of Existing Methods. }Existing \textit{MTL and acoustic feature-based approaches} are highly dependent on textual information. This text-dependency bias may inherently limit the applicability of these methods when semantics are benign but paralinguistic properties, such as intonation and emotion, convey harmful intent. Current \textit{feature fusion methods} focus on combining specific acoustic dimensions, which may miss subtle paralinguistic toxic signals that exist beyond their explicitly extracted dimensions. Some \textit{acoustic feature-based methods} rely on traditional handcrafted features for toxicity detection, which may fail to capture rich dimensions representing harmful intent. DeToxy~\cite{ghosh22b_interspeech} applies Self-Supervised Learning (SSL)~\cite{liu2022audio, gong2022ssast} pre-trained foundation models, but underutilizes their representational capabilities through simple feature extraction without sophisticated architectural design. Moreover, DeToxy focuses solely on textual content analysis, failing to address toxicity that originates from paralinguistic sources.

\noindent\textbf{Evaluation and Reproducibility Limitations. }The benchmarking practices in this field are incomplete and inconsistent. Existing works primarily compare against their own baselines, lacking broader evaluations. The lack of open-source code further hinders reproducibility and collaboration in the field. These limitations collectively create substantial obstacles to advancing research in paralinguistic-aware toxic speech detection, highlighting the urgent need for comprehensive datasets, transparent methodologies, and reproducible evaluation frameworks.             

In this work, we address these limitations from both data and methodological perspectives:

\noindent (1) We develop and open-source ToxiAlert-Bench, the first large-scale toxic speech dataset specifically designed for paralinguistic-aware detection, comprising over 60 hours of annotated audio clips. It features comprehensive toxicity source annotations, including four distinct categories: safe for both textual and paralinguistic sources, textually toxic but paralinguistically safe, textually safe but paralinguistically toxic, and toxic for both sources. The dataset encompasses seven toxic categories and a safe category, with twenty fine-grained toxic labels. We systematically document and open-source the complete dataset construction pipeline, enabling reproducible research and facilitating future dataset-building studies in this domain.

\noindent (2) We propose a novel dual-head neural network architecture built upon pre-trained SSL foundation models for robust toxic speech detection. Our model leverages large-scale pre-trained representations to capture both semantic and paralinguistic features effectively. The architecture incorporates two specialized classification heads with a multi-stage training strategy. To address data imbalance challenges, we integrate class-balanced sampling and weighted loss functions. Extensive experimental results, including benchmarking and ablation results, validate the effectiveness of our model architecture and training strategies.

Our contributions are summarized as follows:
\begin{enumerate}
    \item We fill the gap in the research domain of toxic speech detection, with the documentation and open source ToxiAlert-Bench, a comprehensive paralinguistic-aware toxic speech dataset. 
    \item We design a dual-head speech detection framework, employing a multi-stage training strategy with class-balanced sampling, weighted loss functions, and sequential head-specific training followed by joint fine-tuning. 
    \item Through comprehensive benchmarking against established baselines, including DeToxy and state-of-the-art (SOTA) multimodal large language models (MLLM), our approach demonstrates significant improvements, achieving a 21.1\% relative improvement in Macro-F1 and a 13.0\% relative gain in accuracy over the strongest baseline.
\end{enumerate}

\section{Related Work}
\subsection{Textual Content Moderation}
Conventional toxic speech detection often ignores the non-verbal properties of speech signals. Some early content moderation (CM) methods heavily depend on manual examination, which is costly and non-scalable. Platforms commonly employ automated CM to ensure that content aligns with behavioral standards by removing inappropriate posts and spam. Most moderators on social platforms utilize conventionally text-based frameworks~\cite{lin2022toxic, nada2023lightweight, koratana2018toxic}. They identify whether a post or comment contains toxic information by analyzing the textual features.

\subsection{Audio-Based Toxic Speech Detection}
Studies in this area can be divided into three categories. \textbf{General Acoustic Features:} Yousefi~et~al.~\cite{Yousefi2021Audio} propose a self-attentive Convolutional Neural Networks framework to detect audio-based toxic language. DeToxy~\cite{ghosh22b_interspeech} proposes to use acoustic features (F-Bank and wav2vec2.0) for classification. Liu~et~al.~\cite{liu2024enhancing} propose a cross-modal learning to incorporate semantic information of text into audio feature representatives, facilitating speech toxicity classification only requiring audio. \textbf{Multi-Task Learning with text information:} Nada~et~al.~\cite{nada2023lightweight} applies an Automatic Speech Recognition (ASR) task to assist toxicity detection. Liu~et~al.~\cite{liu2024enhancing2} predict the toxicity labels of a speech signal with the assistance of text information alignment. Nandwana et al.\cite{nandwana2024voice} utilize multi-task learning to predict the toxicity of speech with the assistance of an auxiliary Audio Keyword detection task. \textbf{Feature Fusion:} Lin~et~al.~\cite{lin2022toxic} explore the relationship between speech emotion and toxic speech and propose a framework combining speech emotion recognition (SER) and audio-based CM models. Rana~et~al.~\cite{rana2022emotion} combines acoustic features representing emotion and text features for hate speech detection. Attentive fusion~\cite{mandal2024attentive} fuses audio and text representation for hate speech identification.

\section{ToxiAlert-Bench}

\begin{figure*}[t]
\centering
\includegraphics[width=1\textwidth]{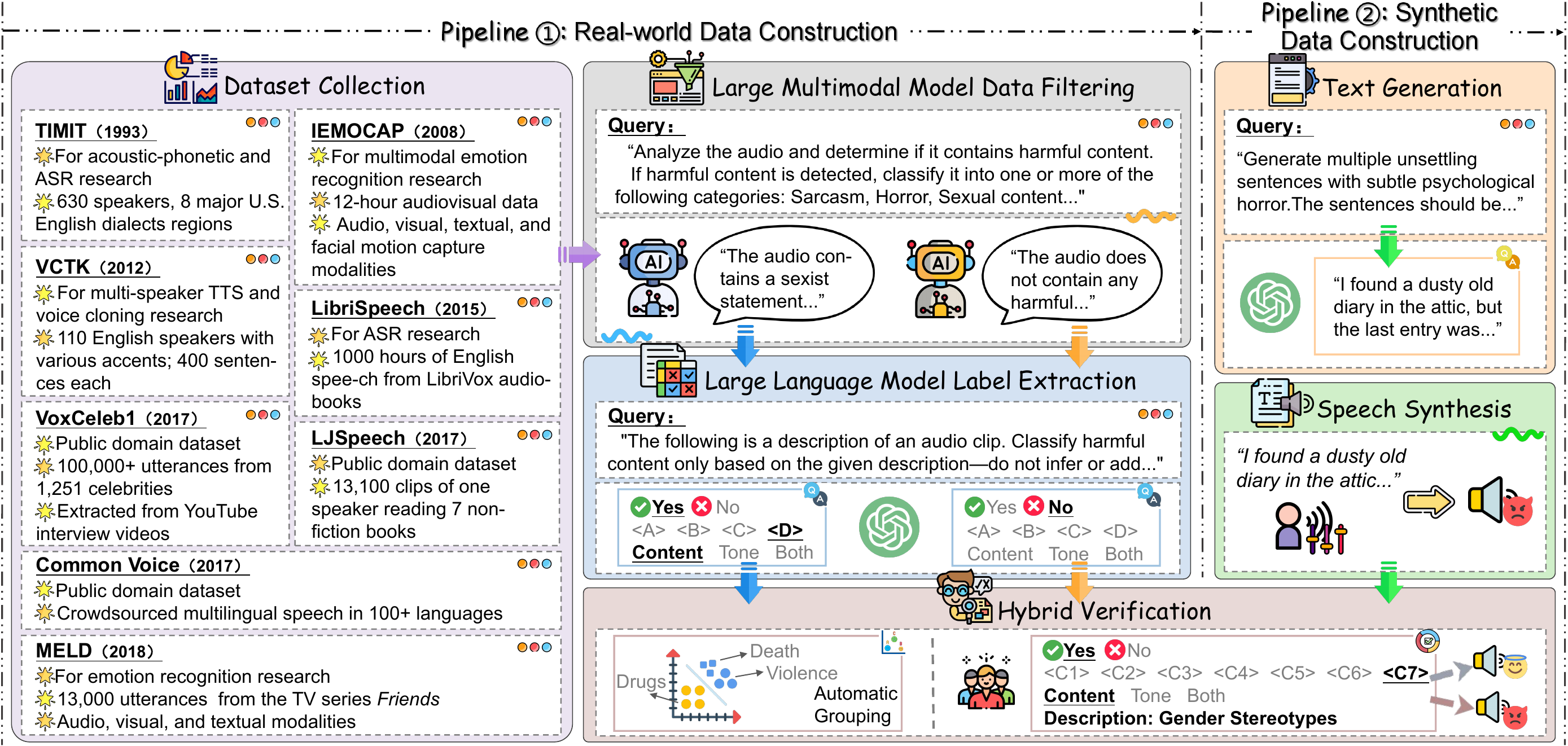} % Reduce the figure size so that it is slightly narrower than the column.
\caption{Overview of the ToxiAlert-Bench dataset construction framework. Pipeline1 (left) illustrates the collection and annotation process for bonafide real-world audio data. Pipeline2 (right) depicts the generation of synthetic toxic speech.}
\label{method1}
\end{figure*}

\subsection{Overview}
ToxiAlert-Bench is a comprehensive English toxic speech dataset comprising 32,561 audio samples totaling 60.82 hours. It uniquely combines both real-world and synthesized audio, including 19,745 samples from established speech corpora and 12,816 samples specifically synthesized for toxicity analysis. To facilitate rigorous experimentation, ToxiAlert-Bench is split into training, validation, and test sets using a 7:1:2 ratio. Each audio sample is annotated with three key attributes: (1) toxicity classification across 7 major categories—\textbf{C1: Sarcasm, C2: Horror, C3: Sexual, C4: Mental Health \& Risk Behavior}(Mental \& Risk), \textbf{C5: Political \& Ideological Sensitivity}(Ideology), \textbf{C6: Violence \& Harm, and C7: Discrimination} plus a \textbf{Safe} category; (2) toxicity source identification, distinguishing between textual-only toxic, paralinguistic-only toxic, both textual and paralinguistic toxic, or safe, and (3) fine-grained categorization, using 20 specific toxicity labels within the taxonomy. Notably, our dataset focuses on paralinguistic toxicity detection—6,728 samples exhibit toxicity solely through paralinguistic cues, addressing a critical gap in existing datasets that primarily focus on textual toxicity.

Compared to the existing DeToxy-B dataset~\cite{ghosh22b_interspeech}, ToxiAlert-Bench demonstrates advantages in both scale and annotation comprehensiveness. While DeToxy-B contains 20,271 utterances (24.4 hours), ToxiAlert-Bench includes 60\% more samples and 150\% more total duration. Importantly, DeToxy-B defines toxicity purely based on textual content, with all 5,077 toxic samples labeled through text-only analysis. In contrast, ToxiAlert-Bench introduces a novel toxicity source annotation framework, explicitly labeling samples as textual-only toxic (6,953), paralinguistic-only toxic (6,728), both textual and paralinguistic toxic (2,551), and safe content (16,329). This granular source-based annotation enables researchers to develop models capable of identifying paralinguistic toxicity. Furthermore, while DeToxy-B does not provide detailed toxicity type labels, our dataset provides comprehensive coverage with 7 major toxic categories and 20 fine-grained labels, supporting more nuanced toxic speech research. Please refer to Appendix A for more details on ToxiAlert-Bench.

\subsection{Dataset Construction Framework}
\noindent\textbf{Bonafide Data Sources. }We collect bonafide speech samples from eight datasets widely accepted in the domain of speech-related studies. They are: (1) TIMIT~\cite{garofolo1993darpa}, (2) IEMOCAP~\cite{busso2008iemocap}, (3) VCTK~\cite{veaux2017cstr}, (4) LibriSpeech~\cite{panayotov2015librispeech}, (5) VoxCeleb1~\cite{nagrani2017voxceleb}, (6) LJSpeech-1.1~\cite{ljspeech17}, (7) CommonVoice~\cite{ardila2019common}, and (8) MELD~\cite{poria2018meld}. For each one, we collect both non-toxic and toxic samples, and these samples are further categorized with a multi-stage annotation pipeline.

\noindent\textbf{Annotation Pipeline for Bonafide Data.} 
Our annotation pipeline for bonafide data follows a systematic multi-stage procedure, as shown in Figure~\ref{method1}. We first employ two large multimodal models, Gemini-1.5-Flash~\cite{geminiteam2024gemini15} and R1-AQA~\cite{li2025reinforcement}\footnote{R1-AQA is based on Qwen2-Audio-7B-Instruct~\cite{Qwen2-Audio}, optimized through reinforcement learning (RL), achieving SOTA performance on the MMAU benchmark~\cite{sakshi2024mmau} with only 38k post-training samples.}, for initial data filtering and preliminary toxicity assessment. Each audio sample is processed using a structured query that determines whether it contains harmful content(see Appendix B for the complete prompt and question-answer pairs). During this stage, both multimodal models analyze each audio sample and its description along three dimensions: toxicity (toxic/non-toxic), major toxic category, and source. When harmful content is detected, the sample is assigned to one of four major toxic categories that serve as our coarse taxonomy: Sarcasm, Horror, Sexual Content, and Other Harmful Content (Class D). This taxonomy is based on our observation that the first three categories are the only ones that consistently show paralinguistic-only toxicity, such as sarcastic tone. Other harmful behaviors lack stable patterns, and defining additional categories would introduce manual bias. To avoid introducing such bias, all remaining harmful instances are grouped into a single catch-all class.

In the next phase, GPT-4o is used to extract structured label suggestions regarding toxicity, coarse category, and source. These suggestions are combined with the multimodal model outputs for a consistency check. When both models reach consensus on all three aspects, the corresponding labels are automatically assigned and subsequently verified through light human review. If any disagreement arises, the sample is forwarded to re-annotation for validation. During this process, the annotator also provides a free-form description for each sample categorized as Class D, characterizing their specific toxic traits. These descriptions are subsequently clustered using unsupervised algorithms\cite{likas2003global} to reveal latent fine-grained toxicity types, and the annotator corrects cluster assignments to ensure semantic coherence (see Appendix A for clustering details). This hybrid clustering approach results in 20 distinct toxic labels plus one safe category. These labels are organized into 7 major toxic classes with hierarchical grouping accomplished by the annotator.

\noindent\textbf{Synthesized Data Construction.}
To enhance diversity and ensure comprehensive coverage of paralinguistic toxicity patterns, we implement a synthetic data generation pipeline, as illustrated in the right portion of Figure~\ref{method1}. We utilize the text-to-speech (TTS) method for synthesizing data~\cite{eskimez2024e2, chen2024f5, anastassiou2024seed}.

Our synthesis begins with GPT-4o generating emotionally charged sentences across toxic categories like psychological horror and subtle sexual tension. Carefully designed prompts (see Appendix B) emphasize subtle and context-dependent expressions of toxicity while avoiding explicit language. GPT-4o's strong safety mechanisms ensure all generated sentences are non-toxic on the textual level, allowing paralinguistic cues to be the sole carriers of toxicity.

Following text generation, we use DubbingX~\cite{dubbingX2025} to synthesize audio. The TTS engine features extensive character personality configurations, enabling the speech generation with distinct vocal styles by simply providing the input text and selecting a specific character role. We strategically select personas likely to produce speech with different toxic paralinguistic characteristics. The synthesis process enables us to produce naturalistic speech with varying paralinguistic features, ensuring that the toxicity of the resulting audio manifests itself through non-textual cues, such as intonation, rhythm, and emotional expression. 

\noindent\textbf{Expert Proofreading.}
After constructing both bonafide and synthetic data, human annotators assess each sample for undergoes detailed evaluation: (1) whether the content contains harmful elements, (2) the source of toxicity (textual content, paralinguistic cues, or both), and (3) the specific toxic category. To ensure annotation quality, two domain experts independently labeled all samples, reaching Cohen’s $\kappa$ = 0.82~\cite{cohen1960coefficient}; disagreements were resolved through discussion.

\begin{figure*}[t] 
\centering
\includegraphics[width=1\textwidth]{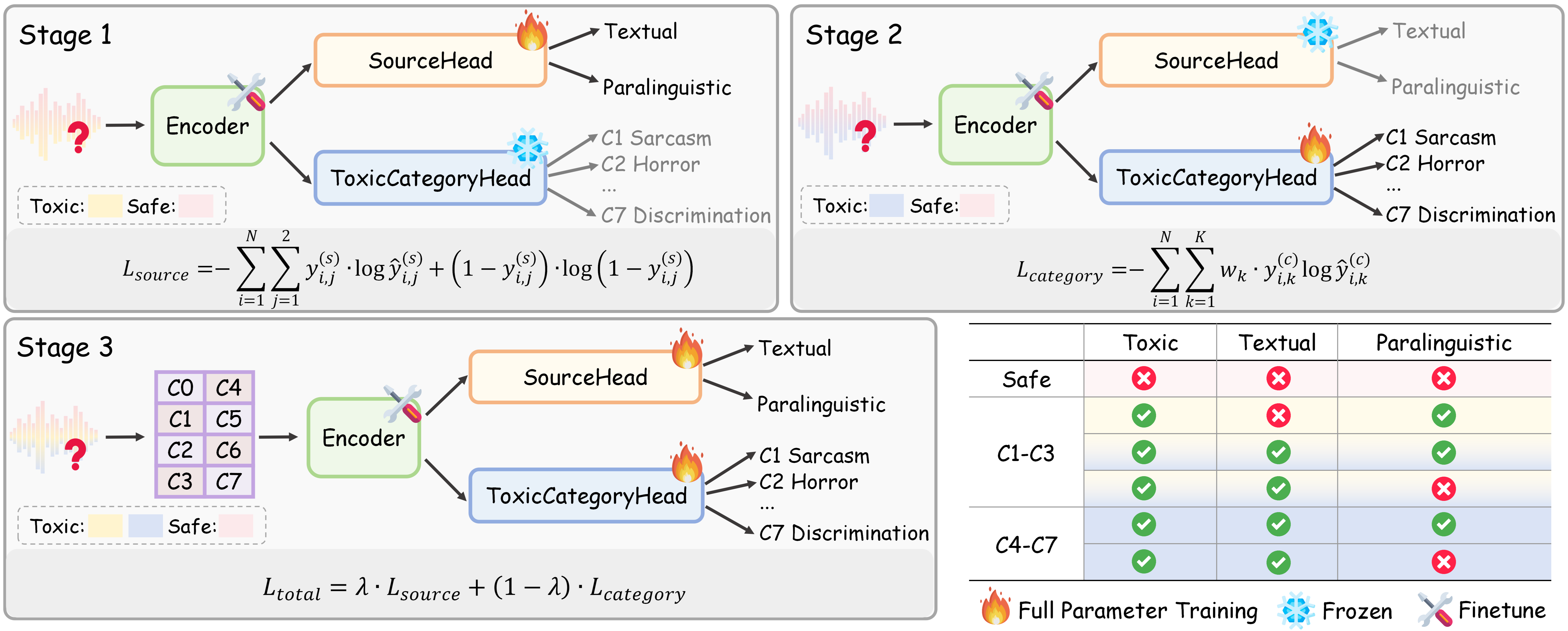} % Reduce the figure size so that it is slightly narrower than the column.
\caption{Overview of the ToxiAlert training framework. Multi-Stage Training Strategy: Stage 1 trains the source head to detect toxicity sources; Stage 2 trains the category head for toxicity classification; Stage 3 jointly fine-tunes both heads.}
\label{method2}
\end{figure*}

\section{ToxiAlert}
To address the limitations of existing approaches, we propose ToxiAlert, a unified detection model designed to identify toxic speech where toxicity may arise from textual content, paralinguistic cues, or their combination. 

\subsection{Design Principles}
We desire to utilize the multi-dimensional information of speech signals to perform toxicity detection. Inspired by recent advancements in the domain of deepfake detection~\cite{tak2022automaticspeakerverificationspoofing}, SOTA methods explore the use of self-supervised learning to obtain better representations trained on diverse speech data and other tasks with only bonafide samples for the purpose of generalization improvement. The pre-trained SSL model, combined with a classifier, is then finetuned with the downstream task dataset, achieving leading performance. Specifically, we adopt Wav2Vec 2.0~\cite{baevski2020wav2vec20frameworkselfsupervised} as the speech encoder $f_\theta: \mathcal{X} \rightarrow \mathbb{R}^d$, where input audio waveform $x \in \mathcal{X}$ is mapped to a latent representation:

\begin{equation}
    \mathbf{h} = f_\theta(x)
\end{equation}

This representation $\mathbf{h} \in \mathbb{R}^{T \times d}$ is passed through two classification heads:
\begin{itemize}
    \item \textbf{Source Head} ($g_\phi^{(s)}$): This is a multi-label classification head designed for toxicity source identification. It predicts whether the toxicity in the audio arises from \textit{textual content}, \textit{paralinguistic cues}, or \textit{both}.
    \begin{equation}
        \hat{\mathbf{y}}^{(s)} = \sigma(g_\phi^{(s)}(\mathbf{h}))
    \end{equation}
    where $\hat{\mathbf{y}}^{(s)} \in [0,1]^2$ represents the independent probabilities assigned to the two binary toxicity sources—textual and paralinguistic. $\sigma(\cdot)$ represents the element-wise \textit{sigmoid} activation function.

    \item \textbf{Category Head} ($g_\phi^{(c)}$): This is a multi-class classification head designed for toxic category classification. It determines the specific type of toxicity in the input audio.
    \begin{equation}
        \hat{\mathbf{y}}^{(c)} = \text{softmax}(g_\phi^{(c)}(\mathbf{h}))
    \end{equation}
    where $\hat{\mathbf{y}}^{(c)} \in [0, 1]^K$ the softmax-normalized likelihoods across $K=8$ mutually exclusive classes, including seven toxic categories and one safe category.
\end{itemize}

\subsection{Multi-Stage Training Strategy}
Our training method employs a multi-stage approach designed to optimize both task-specific performance and inter-task coordination, as illustrated in Figure~\ref{method2}. Let $\mathcal{D}^{(s)}$, $\mathcal{D}^{(c)}$, and $\mathcal{D}^{(full)}$ represent the datasets used in each stage.

\noindent\textbf{Stage 1: Source Head Training.} We first train the source head $g_\phi^{(s)}$, freezing the category head $g_\phi^{(c)}$. This stage focuses on learning to distinguish between textual and paralinguistic toxicity sources. Therefore, the source-training dataset $\mathcal{D}^{(s)}$ is constructed exclusively from the categories Sarcasm, Horror, and Sexual (C1–C3 in Figure~\ref{method2}), since these three categories contain examples across the complete range of source conditions. To achieve class balance, we supplement the toxic samples with safe samples equivalent to approximately 1/3 of the total toxic sample count from these three categories. The objective is a binary cross-entropy loss:

\begin{equation}
    \scalebox{0.88}{$\displaystyle
    \mathcal{L}_{\text{source}} = - \sum_{i=1}^{N} \sum_{j=1}^{2} \left[ y_{i,j}^{(s)} \log \hat{y}_{i,j}^{(s)} + (1 - y_{i,j}^{(s)}) \log(1 - \hat{y}_{i,j}^{(s)}) \right]
    $}
\end{equation}

where $y_{i,j}^{(s)} \in \{0, 1\}$ are the ground-truth binary labels.

\noindent\textbf{Stage 2: Category Head Training.} In the second stage, we freeze the source head $g_\phi^{(s)}$ and train the category head $g_\phi^{(c)}$ on $\mathcal{D}^{(c)}$. The dataset is derived from all toxic categories (C1–C7 in Figure~\ref{method2}) and includes textually toxic but paralinguistically safe samples, allowing the model to focus on textual discrimination. We add safe speech samples of approximately 1/7 of the toxic subset to maintain balance. The training minimizes a weighted cross-entropy loss:

\begin{equation}
    \mathcal{L}_{\text{category}} = - \sum_{i=1}^{N} \sum_{k=1}^{K} w_k \cdot y_{i,k}^{(c)} \log \hat{y}_{i,k}^{(c)}
\end{equation}

where $w_k$ is the inverse frequency of class $k$ for balancing, and $y_{i,j}^{(c)} \in \{0, 1\}$ are one-hot toxicity category labels.

\noindent\textbf{Stage 3: Joint Fine-tuning.} The final stage performs end-to-end joint training of both heads using the complete dataset $\mathcal{D}^{(full)}$, and a composite objective function is optimized:

\begin{equation}
    \mathcal{L}_{\text{total}} = \lambda \cdot \mathcal{L}_{\text{source}} + (1 - \lambda) \cdot \mathcal{L}_{\text{category}}
\end{equation}

where $\lambda = 0.2$ to reflect the relatively auxiliary nature of the source task.

To ensure training stability and mitigate label imbalance, we employ a class-balanced sampler that selects $m$ samples per category for every batch of size $B = m \cdot K$. In our experiments, we use $m = 3$, resulting in $B = 24$.

\setlength{\tabcolsep}{1.5mm} % 压缩列间距
\begin{table*}[ht]
\centering
\footnotesize
\begin{tabular}{l|ccccccc|cc|c}
\toprule
\textbf{Model} 
& \textbf{Sarcasm} 
& \textbf{Horror} 
& \textbf{Sexual} 
& \makecell{\textbf{Mental} \\ \textbf{\& Risk}} 
& \textbf{Ideology} 
& \makecell{\textbf{Violence} \\ \textbf{\& Harm}} 
& \textbf{Discrim.} 
& \makecell{\textbf{ACC}} 
& \makecell{\textbf{Macro-F1}}
& \makecell{\textbf{Binary} \\ \textbf{ACC}} \\
\midrule
DeToxy & - & - & - & - & - & - & - & - & - & 85.70 \\
YIDUN & - & - & 0.50 & - & 0.50 & 0.65 & - & - & - & 50.49 \\
Qwen2-Audio & 4.42 & 0.00 & 12.21 & 0.00 & 2.51 & 26.83 & 9.73 & 55.15 & 19.24 & 60.41 \\
Gemini-2.5-Flash & 53.00 & 58.89 & 34.32 & 47.15 & 21.61 & 61.64 & 36.19 & 70.84 & 57.55 & 75.38 \\
GPT-4o Audio & 27.08 & 12.22 & 20.17 & 29.27 & 18.09 & 34.88 & 21.01 & 61.89 & 39.91 & 64.52 \\
\midrule
\textbf{ToxiAlert} & \textbf{81.10} & \textbf{90.94} & \textbf{81.85} & \textbf{48.78} & \textbf{52.76} & \textbf{65.95} & \textbf{39.30} & \textbf{80.04} & \textbf{69.69} & \textbf{86.33} \\
\bottomrule
\end{tabular}
\caption{Comparison of ToxiAlert with baselines on ToxiAlert-Bench. We report per-category accuracy across seven toxicity categories. Note that models without category-level predictions leave corresponding entries blank (-).}
\label{tab:toxicity-results}
\end{table*}

\begin{table}[t]
\centering
\footnotesize
\begin{tabular}{ll|cccc|c}
\toprule
\multicolumn{2}{c|}{} & \multicolumn{4}{c|}{\textbf{Label-Level}} & \makecell{\textbf{Sample-} \\ \textbf{Level}} \\
\cmidrule(lr){3-6} \cmidrule(lr){7-7}
\textbf{Model} & \textbf{Type} & ACC & F1 & \makecell{Macro \\ F1} & \makecell{Micro \\ F1} & \makecell{Subset \\ ACC} \\
\midrule
\multirow{2}{*}{Qwen2} 
& Para. & 71.84 & 3.79  & \multirow{2}{*}{19.28} & \multirow{2}{*}{20.72} & \multirow{2}{*}{55.35} \\
& Tex.  & 77.00 & 34.77 &                      &                      &                      \\
\multirow{2}{*}{Gemini} 
& Para. & 69.48 & 19.57 & \multirow{2}{*}{31.11} & \multirow{2}{*}{31.31} & \multirow{2}{*}{52.90} \\
& Tex.  & 77.48 & 42.66 &                      &                      &                      \\
\multirow{2}{*}{GPT-4o} 
& Para. & 71.50 & 0.32  & \multirow{2}{*}{13.81} & \multirow{2}{*}{15.04} & \multirow{2}{*}{53.20} \\
& Tex.  & 75.06 & 27.30 &                      &                      &                      \\
\midrule
\multirow{2}{*}{\textbf{ToxiAlert}} 
& \textbf{Para.} & \textbf{91.18} & \textbf{83.30} & \multirow{2}{*}{\textbf{79.48}} & \multirow{2}{*}{\textbf{79.34}} & \multirow{2}{*}{\textbf{80.21}} \\
& \textbf{Tex.}  & \textbf{86.21} & \textbf{75.66} &                             &                             &                             \\
\bottomrule
\end{tabular}
\caption{Comparison of model performance on the source identification task. Both label-level and sample-level results are reported.}
\label{tab:toxicity source}
\end{table}

\section{Experiments}
\subsection{Settings}
\noindent\textbf{Baselines. }We compare ToxiAlert with several SOTA open-source and commercial systems. DeToxy and NetEase Yidun Audio Moderation API (YIDUN)~\cite{NetEaseYidun} are specifically built for toxic speech detection. DeToxy is an open-source toxicity classifier, while YIDUN is a commercial platform supporting real-time moderation in multiple languages. In contrast, Qwen2-Audio, GPT-4o Audio, and Gemini-2.5-Flash~\cite{comanici2025gemini} are general-purpose MLLMs not explicitly trained for toxicity detection, these models have demonstrated strong capabilities in speech comprehension and multimodal reasoning, due to their large parameter scale and training on vast datasets. 

\noindent\textbf{Evaluation Setup. }We train ToxiAlert and DeToxy on the ToxiAlert-Bench training set and directly evaluate other baselines. Toxicity classification performance is assessed at both the category level (7 toxic categories) and the label level (20 fine-grained labels). In addition to overall performance, we emphasize a challenging subset of the benchmark where the toxicity is conveyed solely through paralinguistic cues. This setting remains underexplored in prior work, yet it is highly relevant for real-world applications. For generalization evaluation, all models are tested on the DeToxy-B test set\footnote{After excluding CMU-MOSEI, CMU-MOSI, MSP-Improv, MSP-Podcast, Social-IQ, and SwitchBoard due to their discontinued open access, the test set contains 2,035 samples.}. Evaluation prompts are detailed in Appendix B.

\noindent\textbf{Metrics. }To comprehensively evaluate model performance, we adopt metrics from two tasks: (1) Toxicity category classification, reporting overall accuracy, per-category accuracy, and Macro-F1 to capture both global and class-specific performance. For binary classifiers like DeToxy, we compute binary accuracy by merging all toxic classes for fair comparison. (2) Toxicity source identification is formulated as a multi-label task, evaluated with label-level metrics—accuracy, F1 score, Macro-F1, and Micro-F1. We also report subset accuracy at the sample-level, which measures the percentage of samples with all labels predicted correctly.

\noindent\textbf{Implementation Details. }We adopt wav2vec2-large-960h as the audio encoder, followed by three fully connected layers for toxicity classification and source identification. All audio samples are resampled to 16kHz, and truncated to a maximum length of 25 seconds. All experiments are conducted on NVIDIA A100 GPUs using PyTorch.

\begin{figure}[t]
\centering
\includegraphics[width=1\columnwidth]{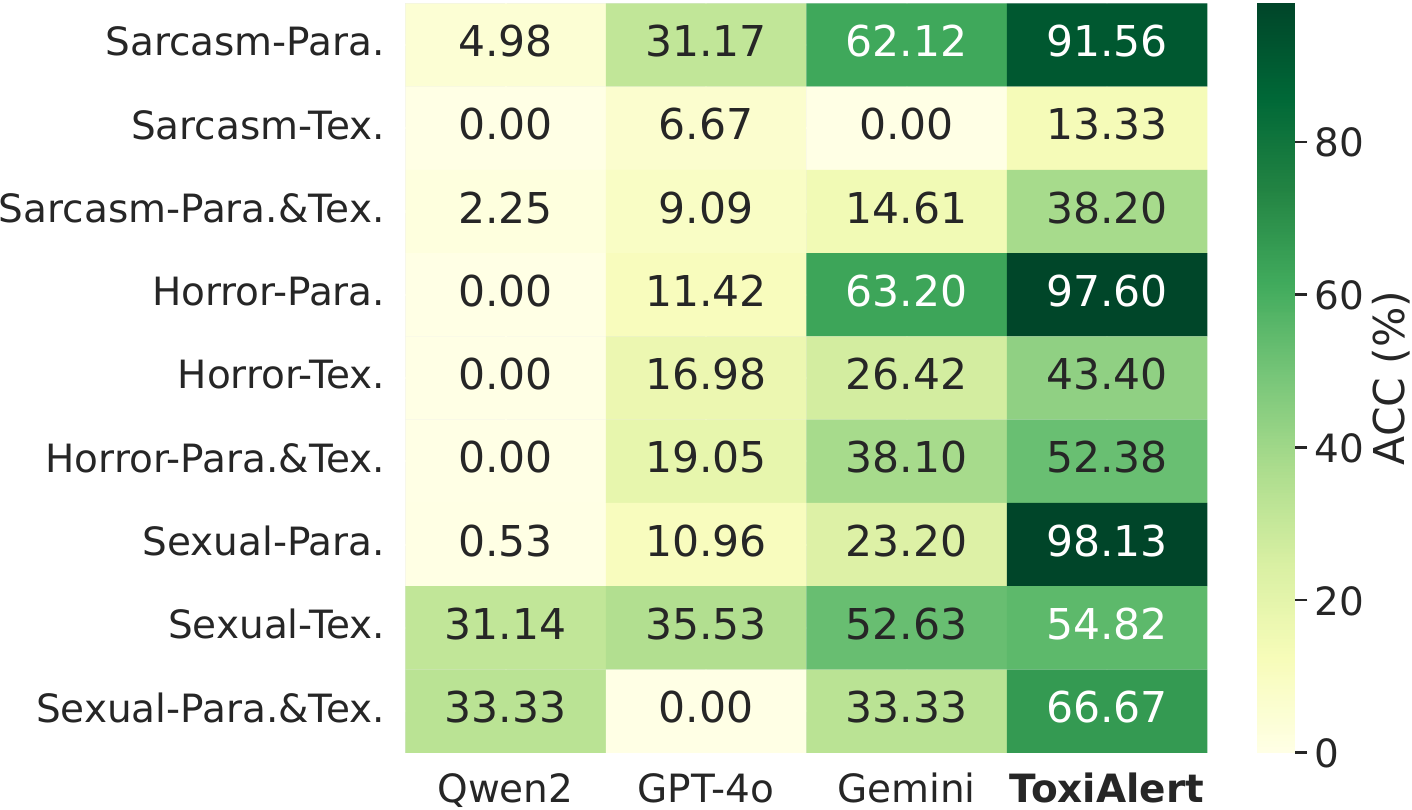} % Reduce the figure size so that it is slightly narrower than the column. Don't use precise values for figure width.This setup will avoid overfull boxes.
\caption{Performance comparison on source-specific toxicity detection across three toxicity types and three source settings. ACC denotes per-class accuracy.}
\label{exp2}
\end{figure}

\begin{figure*}[t]
\centering
\includegraphics[width=1\textwidth]{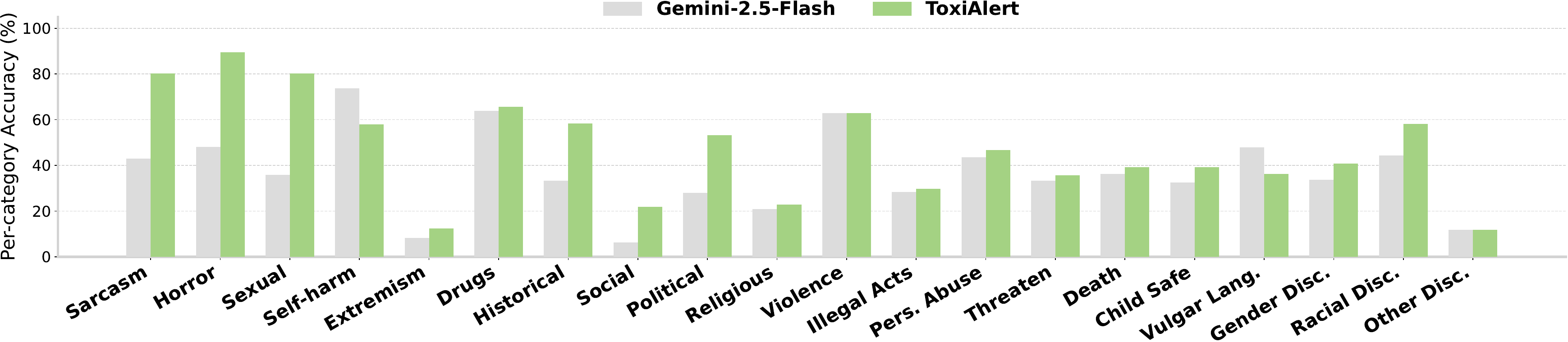} % Reduce the figure size so that it is slightly narrower than the column.
\caption{Fine-grained comparison of ToxiAlert and Gemini-2.5-Flash on ToxiAlert-Bench. We report per-category accuracy across twenty fine-grained toxicity labels, spanning all seven major toxicity categories.}
\label{exp1}
\end{figure*}

\subsection{Toxic Speech Classification}
\noindent\textbf{Category Level:} 
As shown in Table~\ref{tab:toxicity-results}, ToxiAlert consistently achieves the best overall performance across all 7 toxicity categories. Compared with the strongest baseline, Gemini-2.5-Flash, it improves Macro-F1 by 21.1\% and overall accuracy by 13.0\%. While DeToxy reports high binary accuracy, but being a binary classifier, it lacks the capacity to distinguish between toxicity types, making it less suitable for fine-grained moderation. In contrast, ToxiAlert delivers both higher binary accuracy and comprehensive multi-class prediction.

\noindent\textbf{Label Level:} 
Given that Gemini-2.5-Flash achieves the best performance at the category level, we adopt it as the baseline for assessing fine-grained classification capabilities. As shown in Figure~\ref{exp1}, ToxiAlert outperforms Gemini-2.5-Flash on the majority of labels. These improvements suggest that ToxiAlert is better equipped to distinguish subtle differences among overlapping or co-occurring toxic behaviors. 

\subsection{Source-Specific Toxicity Detection}
We further assess model performance under varied source conditions of toxic expression. Specifically, we focus on three challenging categories—Sarcasm, Horror, and Sexual, and evaluate classification accuracy when the toxic signal is conveyed through paralinguistic cues (Para.), textual content (Tex.), or both (Para.\&Tex.). Results are presented in Figure~\ref{exp2}. ToxiAlert consistently outperforms all baselines across all categories and source types. In cases where toxic intent is expressed exclusively through Para., ToxiAlert achieves 91.56\% on Sarcasm, 97.60\% on Horror, and 98.13\% on Sexual. In contrast, baselines show notable performance degradation, as they typically overlook non-verbal signals during training or inference. 

\begin{table}[t]
\centering
\footnotesize

\begin{tabular}{l|cc|c}
\toprule
\textbf{Model} & \makecell{Balanced ACC} & \makecell{F1-Binary} & \makecell{Toxic ACC} \\
\midrule
DeToxy        & 66.95 & 50.33 & 67.78 \\
YIDUN         & 49.97 & 0.40  & 0.20  \\
Qwen2-Audio         & 52.91 & 12.50 & 6.88  \\
Gemini-2.5-Flash        & 59.89 & 37.17 & 29.47 \\
GPT-4o Audio        & 69.20 & 54.32 & 48.51 \\
\midrule
\textbf{ToxiAlert} & \textbf{72.29} & \textbf{55.83} & \textbf{80.94} \\
\bottomrule
\end{tabular}
\caption{Comparison of model generalization performance. Balanced ACC mitigates the effect of class imbalance; Toxic ACC is the accuracy on toxic samples.}
\label{tab:toxicity-general}
\end{table}

\begin{table}[t]
\centering
\footnotesize
\setlength{\tabcolsep}{0.9mm} % 稍微增加间距
\begin{tabular}{@{}l|ccc|cc@{}}
\toprule
\multirow{3}{*}{\textbf{Model}} 
& \multicolumn{3}{c|}{\textbf{Toxic Cls.}} 
& \multicolumn{2}{c}{\textbf{Source ID}} \\
\cmidrule(lr){2-4} \cmidrule(lr){5-6}
& \makecell[c]{ACC} & \makecell[c]{Macro-F1} & \makecell[c]{Binary\\ACC} & \makecell[c]{Macro-F1} & \makecell[c]{Subset\\ACC} \\
\midrule
w/o SourceHead     & 75.04 & 66.01 & 81.67 & --    & --    \\
w/o Multi-stage   & 78.25 & 68.79 & 84.72 & 78.35 & 77.80 \\
w/o Sampler  & 78.34 & 68.00 & 85.47 & 79.05 & 79.51 \\
\midrule
\textbf{ToxiAlert} & \textbf{80.04} & \textbf{69.69} & \textbf{86.33} & \textbf{79.48} & \textbf{80.21} \\
\bottomrule
\end{tabular}
\caption{Ablation study on the effectiveness of ToxiAlert components. Performance is shown for toxicity category classification (Toxic Cls.) and toxicity source identification (Source ID).}
\label{tab:ablation}
\end{table}

\subsection{Toxicity Source Identification}
We assess the model’s ability to identify toxicity sources—textual or paralinguistic. As shown in Table~\ref{tab:toxicity source}, ToxiAlert consistently outperforms all baselines. For paralinguistic cues, which are inherently subtle and challenging to detect, ToxiAlert achieves an accuracy of 91.18\% and an F1 score of 83.30\%, significantly surpassing all competing models. For textual content sources, it also delivers strong results with 86.21\% accuracy and 75.66\% F1 score. Moreover, ToxiAlert achieves the highest overall performance on Subset Accuracy, improving over the strongest baseline by 44.9\%, underscoring its robustness in capturing both explicit and implicit forms of toxic expression.

\subsection{Generalization Evaluation}
To assess the generalization ability of ToxiAlert, we evaluate it on DeToxy-B. For fair comparison, we train a binary version of ToxiAlert using the training set of ToxiAlert-Bench and evaluate it directly on the DeToxy-B test set without any additional fine-tuning. As shown in Table~\ref{tab:toxicity-general}, ToxiAlert surpasses the best-performing baseline, GPT-4o Audio, by 4.5\% in balanced accuracy, 2.8\% in F1 score, and 32.9\% in accuracy on toxic samples. These results demonstrate that ToxiAlert generalizes effectively to out-of-distribution data.

\subsection{Ablation Study}
To investigate the impact of each core component in ToxiAlert, we conduct an ablation study focusing on three core modules: the dual-head architecture, the multi-stage training strategy, and the class-balanced sampler. As shown in Table~\ref{tab:ablation}, removing any of these components results in a noticeable performance decline. 

Removing the source head leads to a significant drop in classification accuracy and Macro-F1, which drops to 75.04\% and 66.01\%, respectively. Moreover, the model is no longer capable of performing source identification, highlighting the necessity of this joint modeling approach. Removing the multi-stage training strategy results in consistent degradation across metrics, with overall accuracy reduced to 78.25\% and subset accuracy to 77.80\%. These results underscore the importance of progressive training in improving convergence and generalization. Finally, without the class-balanced sampler, performance in both tasks degrades. This result highlights the importance of structured sampling in improving model stability in multi-class classification tasks.

\section{Conclusion}
In this work, we propose ToxiAlert, the first paralinguistic-toxic-aware speech toxicity dataset, covering diverse toxicity-source combinations and capable of facilitating the development of a more comprehensive toxic speech detection system. We also present an SSL-based model that predicts toxic/safe labels, major toxicity category, and toxicity source. With a dual-head design and multi-stage training strategy, our model outperforms existing academic methods and commercial MLLM-based solutions.

\section{Acknowledgments}
This paper is supported in part by the Zhejiang Provincial Natural Science Foundation of China under Grant (LD24F020010), the National Natural Science Foundation of China (62472372, 62172359, 62441238, 62072395 and U20A20178), the Key Research and Development Program of Hangzhou City (2024SZD1A27), and the Key R\&D Programme of Zhejiang Province (2025C02264).

\bibliography{aaai2026}

@misc{nada2023lightweight,
      title={Lightweight Toxicity Detection in Spoken Language: A Transformer-based Approach for Edge Devices}, 
      author={Ahlam Husni Abu Nada and Siddique Latif and Junaid Qadir},
      year={2023},
      eprint={2304.11408},
      archivePrefix={arXiv},
      primaryClass={cs.SD},
      url={https://arxiv.org/abs/2304.11408}, 
}

@INPROCEEDINGS{lin2022toxic,
  author={Lin, Wei-Cheng and Emmanouilidou, Dimitra},
  booktitle={2022 30th European Signal Processing Conference (EUSIPCO)}, 
  title={Toxic Speech and Speech Emotions: Investigations of Audio-based Modeling and Intercorrelations}, 
  year={2022},
  volume={},
  number={},
  pages={115-119},
  doi={10.23919/EUSIPCO55093.2022.9909856}}

@article{koratana2018toxic,
  title={Toxic speech detection},
  author={Koratana, Animesh and Hu, Kevin},
  journal={URL: https://web. stanford. edu/class/archive/cs/cs224n/cs224n},
  volume={1194},
  year={2018}
}

@INPROCEEDINGS{Yousefi2021Audio,
  author={Yousefi, Midia and Emmanouilidou, Dimitra},
  booktitle={2021 29th European Signal Processing Conference (EUSIPCO)}, 
  title={Audio-based Toxic Language Classification using Self-attentive Convolutional Neural Network}, 
  year={2021},
  volume={},
  number={},
  pages={11-15},
  doi={10.23919/EUSIPCO54536.2021.9616001}}

@misc{rana2022emotion,
      title={Emotion Based Hate Speech Detection using Multimodal Learning}, 
      author={Aneri Rana and Sonali Jha},
      year={2022},
      eprint={2202.06218},
      archivePrefix={arXiv},
      primaryClass={cs.LG},
      url={https://arxiv.org/abs/2202.06218}, 
}

@inproceedings{ghosh22b_interspeech,
  title     = {DeToxy: A Large-Scale Multimodal Dataset for Toxicity Classification in Spoken Utterances},
  author    = {Sreyan Ghosh and Samden Lepcha and S Sakshi and Rajiv Ratn Shah and Srinivasan Umesh},
  year      = {2022},
  booktitle = {Interspeech 2022},
  pages     = {5185--5189},
  doi       = {10.21437/Interspeech.2022-10752},
  issn      = {2958-1796},
}

@inproceedings{liu2024enhancing,
  title     = {Enhancing Automated Audio Captioning via Large Language Models with Optimized Audio Encoding},
  author    = {Jizhong Liu and Gang Li and Junbo Zhang and Heinrich Dinkel and Yongqing Wang and Zhiyong Yan and Yujun Wang and Bin Wang},
  year      = {2024},
  booktitle = {Interspeech 2024},
  pages     = {1135--1139},
  doi       = {10.21437/Interspeech.2024-65},
  issn      = {2958-1796},
}

@article{liu2024enhancing2,
  title={Enhancing multilingual voice toxicity detection with speech-text alignment},
  author={Liu, Joseph and Nandwana, Mahesh Kumar and Pylkk{\~A}{\c{k}}nen, Janne and Heikinheimo, Hannes and McGuire, Morgan},
  journal={arXiv preprint arXiv:2406.10325},
  year={2024}
}

@INPROCEEDINGS{nandwana2024voice,
  author={Kumar Nandwana, Mahesh and He, Yifan and Liu, Joseph and Yu, Xiao and Shang, Charles and Du Bois, Eloi and McGuire, Morgan and Bhat, Kiran},
  booktitle={ICASSP 2024 - 2024 IEEE International Conference on Acoustics, Speech and Signal Processing (ICASSP)}, 
  title={Voice Toxicity Detection Using Multi-Task Learning}, 
  year={2024},
  volume={},
  number={},
  pages={331-335},
  doi={10.1109/ICASSP48485.2024.10448289}}

@misc{mandal2024attentive,
      title={Attentive Fusion: A Transformer-based Approach to Multimodal Hate Speech Detection}, 
      author={Atanu Mandal and Gargi Roy and Amit Barman and Indranil Dutta and Sudip Kumar Naskar},
      year={2024},
      eprint={2401.10653},
      archivePrefix={arXiv},
      primaryClass={cs.CL},
      url={https://arxiv.org/abs/2401.10653}, 
}

@misc{geminiteam2024gemini15,
      title={Gemini 1.5: Unlocking multimodal understanding across millions of tokens of context}, 
      author={Gemini Team},
      year={2024},
      eprint={2403.05530},
      archivePrefix={arXiv},
      primaryClass={cs.CL},
      url={https://arxiv.org/abs/2403.05530}, 
}

@article{li2025reinforcement,
  title={Reinforcement Learning Outperforms Supervised Fine-Tuning: A Case Study on Audio Question Answering},
  author={Li, Gang and Liu, Jizhong and Dinkel, Heinrich and Niu, Yadong and Zhang, Junbo and Luan, Jian},
  journal={arXiv preprint arXiv:2503.11197},
  year={2025},
  url={https://github.com/xiaomi-research/r1-aqa; https://huggingface.co/mispeech/r1-aqa}
}

@misc{sakshi2024mmau,
      title={MMAU: A Massive Multi-Task Audio Understanding and Reasoning Benchmark}, 
      author={S Sakshi and Utkarsh Tyagi and Sonal Kumar and Ashish Seth and Ramaneswaran Selvakumar and Oriol Nieto and Ramani Duraiswami and Sreyan Ghosh and Dinesh Manocha},
      year={2024},
      eprint={2410.19168},
      archivePrefix={arXiv},
      primaryClass={eess.AS},
      url={https://arxiv.org/abs/2410.19168}, 
}

@article{Qwen2-Audio,
  title={Qwen2-Audio Technical Report},
  author={Chu, Yunfei and Xu, Jin and Yang, Qian and Wei, Haojie and Wei, Xipin and Guo,  Zhifang and Leng, Yichong and Lv, Yuanjun and He, Jinzheng and Lin, Junyang and Zhou, Chang and Zhou, Jingren},
  journal={arXiv preprint arXiv:2407.10759},
  year={2024}
}

@Misc{dubbingX2025,
  url = {https://dubbingx.com/},
  author = {DubbingX},
  title = {DubbingX TTS},
  year = {2025}
}

@misc{tak2022automaticspeakerverificationspoofing,
      title={Automatic speaker verification spoofing and deepfake detection using wav2vec 2.0 and data augmentation}, 
      author={Hemlata Tak and Massimiliano Todisco and Xin Wang and Jee-weon Jung and Junichi Yamagishi and Nicholas Evans},
      year={2022},
      eprint={2202.12233},
      archivePrefix={arXiv},
      primaryClass={eess.AS},
      url={https://arxiv.org/abs/2202.12233} 
}

@misc{baevski2020wav2vec20frameworkselfsupervised,
      title={wav2vec 2.0: A Framework for Self-Supervised Learning of Speech Representations}, 
      author={Alexei Baevski and Henry Zhou and Abdelrahman Mohamed and Michael Auli},
      year={2020},
      eprint={2006.11477},
      archivePrefix={arXiv},
      primaryClass={cs.CL},
      url={https://arxiv.org/abs/2006.11477}, 
}

@article{zeng2024shieldgemma,
  title={Shieldgemma: Generative ai content moderation based on gemma},
  author={Zeng, Wenjun and Liu, Yuchi and Mullins, Ryan and Peran, Ludovic and Fernandez, Joe and Harkous, Hamza and Narasimhan, Karthik and Proud, Drew and Kumar, Piyush and Radharapu, Bhaktipriya and others},
  journal={arXiv preprint arXiv:2407.21772},
  year={2024}
}

@article{inan2023llama,
  title={Llama guard: Llm-based input-output safeguard for human-ai conversations},
  author={Inan, Hakan and Upasani, Kartikeya and Chi, Jianfeng and Rungta, Rashi and Iyer, Krithika and Mao, Yuning and Tontchev, Michael and Hu, Qing and Fuller, Brian and Testuggine, Davide and others},
  journal={arXiv preprint arXiv:2312.06674},
  year={2023}
}

@article{garofolo1993darpa,
  title={DARPA TIMIT acoustic-phonetic continous speech corpus CD-ROM. NIST speech disc 1-1.1},
  author={Garofolo, John S and Lamel, Lori F and Fisher, William M and Fiscus, Jonathan G and Pallett, David S},
  journal={NASA STI/Recon technical report n},
  volume={93},
  pages={27403},
  year={1993}
}

@inproceedings{panayotov2015librispeech,
  title={Librispeech: an asr corpus based on public domain audio books},
  author={Panayotov, Vassil and Chen, Guoguo and Povey, Daniel and Khudanpur, Sanjeev},
  booktitle={2015 IEEE international conference on acoustics, speech and signal processing (ICASSP)},
  pages={5206--5210},
  year={2015},
  organization={IEEE}
}

@article{poria2018meld,
  title={Meld: A multimodal multi-party dataset for emotion recognition in conversations},
  author={Poria, Soujanya and Hazarika, Devamanyu and Majumder, Navonil and Naik, Gautam and Cambria, Erik and Mihalcea, Rada},
  journal={arXiv preprint arXiv:1810.02508},
  year={2018}
}

@article{busso2008iemocap,
  title={IEMOCAP: Interactive emotional dyadic motion capture database},
  author={Busso, Carlos and Bulut, Murtaza and Lee, Chi-Chun and Kazemzadeh, Abe and Mower, Emily and Kim, Samuel and Chang, Jeannette N and Lee, Sungbok and Narayanan, Shrikanth S},
  journal={Language resources and evaluation},
  volume={42},
  number={4},
  pages={335--359},
  year={2008},
  publisher={Springer}
}

@article{nagrani2017voxceleb,
  title={Voxceleb: a large-scale speaker identification dataset},
  author={Nagrani, Arsha and Chung, Joon Son and Zisserman, Andrew},
  journal={arXiv preprint arXiv:1706.08612},
  year={2017}
}

@article{ardila2019common,
  title={Common voice: A massively-multilingual speech corpus},
  author={Ardila, Rosana and Branson, Megan and Davis, Kelly and Henretty, Michael and Kohler, Michael and Meyer, Josh and Morais, Reuben and Saunders, Lindsay and Tyers, Francis M and Weber, Gregor},
  journal={arXiv preprint arXiv:1912.06670},
  year={2019}
}

@article{veaux2017cstr,
  title={CSTR VCTK corpus: English multi-speaker corpus for CSTR voice cloning toolkit},
  author={Veaux, Christophe and Yamagishi, Junichi and MacDonald, Kirsten and others},
  journal={University of Edinburgh. The Centre for Speech Technology Research (CSTR)},
  volume={6},
  pages={15},
  year={2017}
}

@misc{ljspeech17,
  author       = {Keith Ito and Linda Johnson},
  title        = {The LJ Speech Dataset},
  howpublished = {\url{https://keithito.com/LJ-Speech-Dataset/}},
  year         = 2017
}

@article{comanici2025gemini,
  title={Gemini 2.5: Pushing the frontier with advanced reasoning, multimodality, long context, and next generation agentic capabilities},
  author={Comanici, Gheorghe and Bieber, Eric and Schaekermann, Mike and Pasupat, Ice and Sachdeva, Noveen and Dhillon, Inderjit and Blistein, Marcel and Ram, Ori and Zhang, Dan and Rosen, Evan and others},
  journal={arXiv preprint arXiv:2507.06261},
  year={2025}
}

@misc{NetEaseYidun,
  title        = {{NetEase Yidun}: AI-Powered Business Security Platform},
  author       = {{NetEase}},
  howpublished = {\url{https://dun.163.com/}},
  note         = {Accessed: 2025-08-01},
  year         = {2025}
}

@article{garg2023handling,
  title={Handling bias in toxic speech detection: A survey},
  author={Garg, Tanmay and Masud, Sarah and Suresh, Tharun and Chakraborty, Tanmoy},
  journal={ACM Computing Surveys},
  volume={55},
  number={13s},
  pages={1--32},
  year={2023},
  publisher={ACM New York, NY}
}

@inproceedings{fortuna2020toxic,
  title={Toxic, hateful, offensive or abusive? what are we really classifying? an empirical analysis of hate speech datasets},
  author={Fortuna, Paula and Soler, Juan and Wanner, Leo},
  booktitle={Proceedings of the Twelfth Language Resources and Evaluation Conference},
  pages={6786--6794},
  year={2020}
}

@inproceedings{hamilton2014streaming,
  title={Streaming on twitch: fostering participatory communities of play within live mixed media},
  author={Hamilton, William A and Garretson, Oliver and Kerne, Andruid},
  booktitle={Proceedings of the SIGCHI conference on human factors in computing systems},
  pages={1315--1324},
  year={2014}
}

@article{scherer1973voice,
  title={The voice of confidence: Paralinguistic cues and audience evaluation},
  author={Scherer, Klaus R and London, Harvey and Wolf, Jared J},
  journal={Journal of Research in Personality},
  volume={7},
  number={1},
  pages={31--44},
  year={1973},
  publisher={Elsevier}
}

@article{liu2022audio,
  title={Audio self-supervised learning: A survey},
  author={Liu, Shuo and Mallol-Ragolta, Adria and Parada-Cabaleiro, Emilia and Qian, Kun and Jing, Xin and Kathan, Alexander and Hu, Bin and Schuller, Bjoern W},
  journal={Patterns},
  volume={3},
  number={12},
  year={2022},
  publisher={Elsevier}
}

@inproceedings{gong2022ssast,
  title={Ssast: Self-supervised audio spectrogram transformer},
  author={Gong, Yuan and Lai, Cheng-I and Chung, Yu-An and Glass, James},
  booktitle={Proceedings of the AAAI Conference on Artificial Intelligence},
  volume={36},
  number={10},
  pages={10699--10709},
  year={2022}
}

@inproceedings{eskimez2024e2,
  title={E2 tts: Embarrassingly easy fully non-autoregressive zero-shot tts},
  author={Eskimez, Sefik Emre and Wang, Xiaofei and Thakker, Manthan and Li, Canrun and Tsai, Chung-Hsien and Xiao, Zhen and Yang, Hemin and Zhu, Zirun and Tang, Min and Tan, Xu and others},
  booktitle={2024 IEEE Spoken Language Technology Workshop (SLT)},
  pages={682--689},
  year={2024},
  organization={IEEE}
}

@article{chen2024f5,
  title={F5-tts: A fairytaler that fakes fluent and faithful speech with flow matching},
  author={Chen, Yushen and Niu, Zhikang and Ma, Ziyang and Deng, Keqi and Wang, Chunhui and Zhao, Jian and Yu, Kai and Chen, Xie},
  journal={arXiv preprint arXiv:2410.06885},
  year={2024}
}

@article{anastassiou2024seed,
  title={Seed-tts: A family of high-quality versatile speech generation models},
  author={Anastassiou, Philip and Chen, Jiawei and Chen, Jitong and Chen, Yuanzhe and Chen, Zhuo and Chen, Ziyi and Cong, Jian and Deng, Lelai and Ding, Chuang and Gao, Lu and others},
  journal={arXiv preprint arXiv:2406.02430},
  year={2024}
}

@article{likas2003global,
  title={The global k-means clustering algorithm},
  author={Likas, Aristidis and Vlassis, Nikos and Verbeek, Jakob J},
  journal={Pattern recognition},
  volume={36},
  number={2},
  pages={451--461},
  year={2003},
  publisher={Elsevier}
}

@article{cohen1960coefficient,
  title={A coefficient of agreement for nominal scales},
  author={Cohen, Jacob},
  journal={Educational and psychological measurement},
  volume={20},
  number={1},
  pages={37--46},
  year={1960},
  publisher={Sage Publications Sage CA: Thousand Oaks, CA}
}

\clearpage

% 附录
\section{Appendix A: Dataset Details}

\subsection{Overview of ToxiAlert-Bench}

\begin{figure*}[ht]
\centering
\includegraphics[width=1\textwidth]{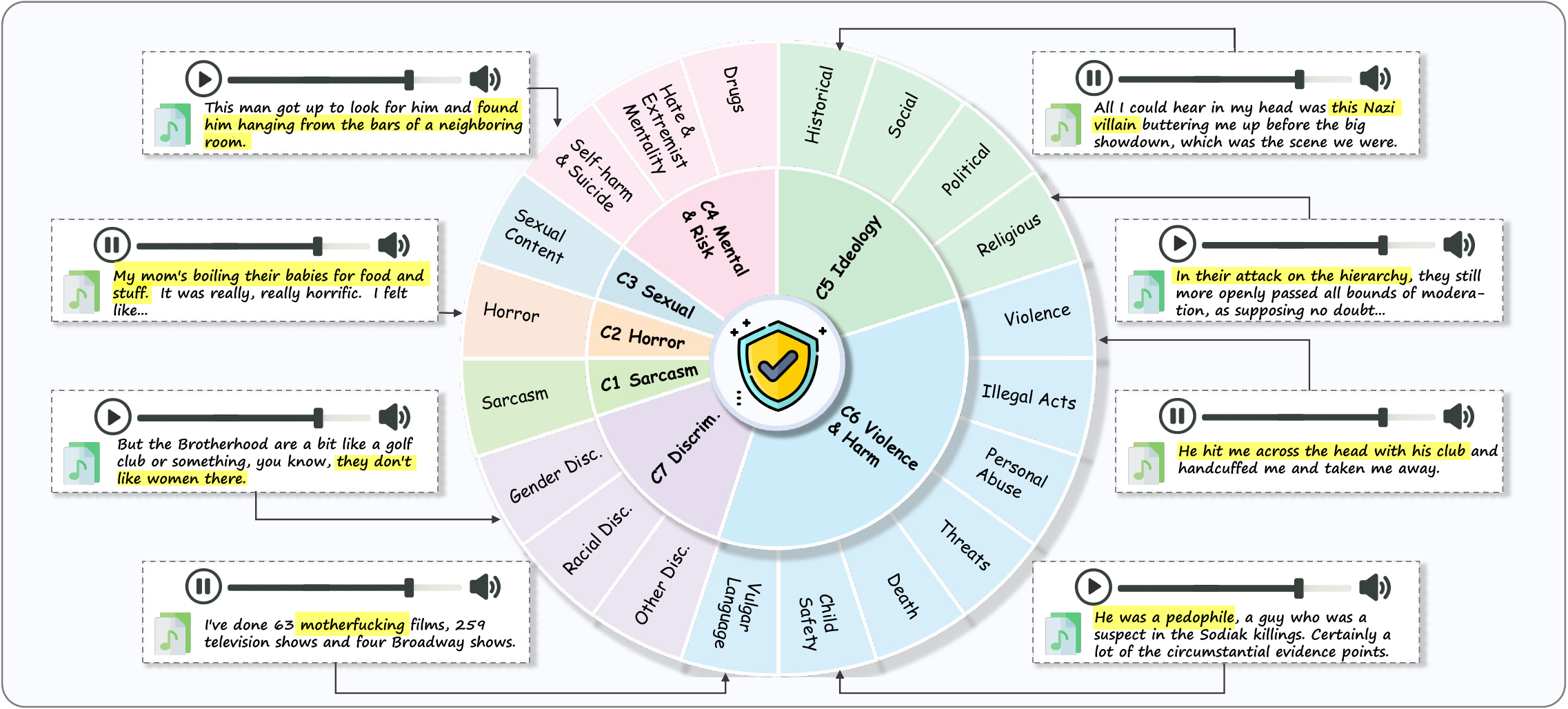}
\caption{Overview of ToxiAlert-Bench taxonomy and examples. The wheel illustrates the 7 coarse-grained toxic categories and their 20 fine-grained subtypes. Each outer example corresponds to a labeled audio utterance, showcasing the diversity of toxicity types captured in the dataset.}
\label{appendix1}
\end{figure*}

Figure~\ref{appendix1} presents a visual overview of ToxiAlert-Bench, illustrating the full taxonomy of toxicity categories along with representative examples. The dataset is hierarchically structured into 7 major toxic categories—Sarcasm, Horror, Sexual, Mental \& Risk, Ideology, Violence \& Harm, and Discrimination. Across these categories, the dataset includes a total of 20 fine-grained labels, capturing nuanced forms of toxic speech such as self-harm, religious content, vulgar language, or drug-related expressions. Specifically, the 20 fine-grained labels include: Sarcasm, Horror, Sexual Content (Sexual), Self-harm \& Suicide (Self-harm), Hate \& Extremist Mentality (Extremism), Drugs, Historical Sensitivity (Historical), Social Sensitivity (Social), Political Sensitivity (Political), Religious Sensitivity (Religious), Violence, Illegal Acts, Personal Abuse, Threats, Death, Child Safety, Vulgar Language, Gender Discrimination, Racial Discrimination, and Other Discrimination.

For each fine-grained label, ToxiAlert-Bench includes curated audio samples accompanied by real or synthesized utterances. The surrounding examples in the figure demonstrate the breadth of toxic content types captured in the dataset, ranging from profane statements and ideological sensitivity to graphic violence. These examples emphasize the diversity and granularity of toxic speech phenomena addressed by ToxiAlert-Bench, and the importance of robust detection models capable of handling both explicit and contextually subtle forms of toxicity.

\subsection{Annotation Format}
\begin{figure}[t]
\centering
\includegraphics[width=0.8\columnwidth]{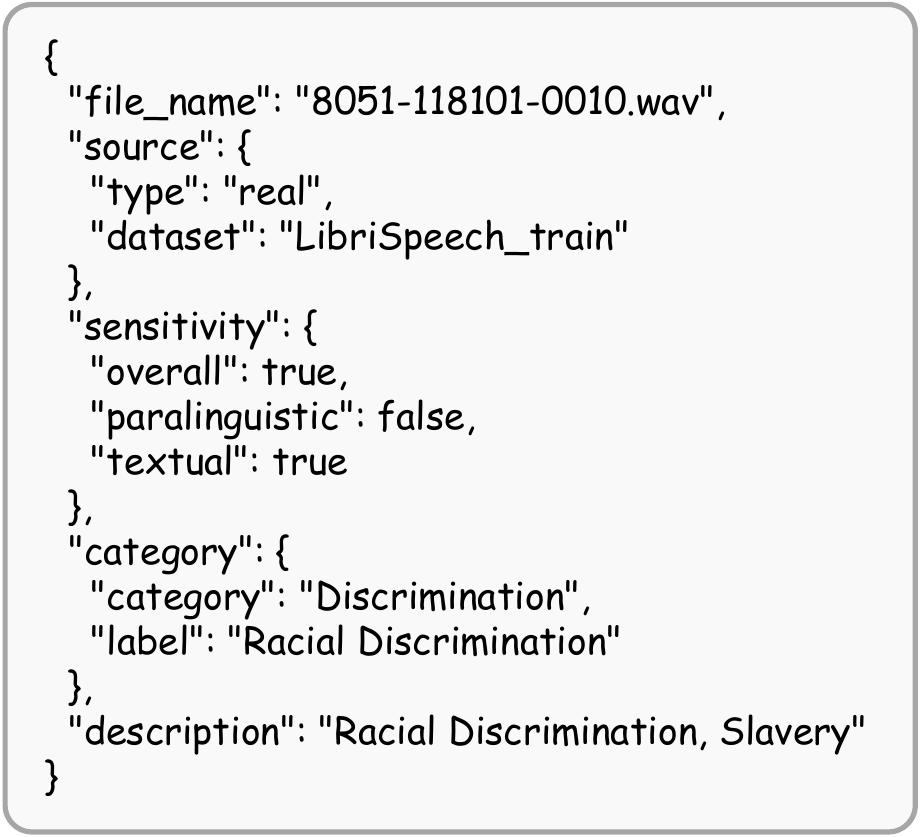} 
\caption{An example annotation from ToxiAlert-Bench in JSON format. Both toxic and non-toxic samples share the same format, with the \texttt{category} and \texttt{label} fields set to \texttt{Safe} when the audio is non-toxic.}
\label{appendix2}
\end{figure}

Each audio sample in ToxiAlert-Bench is annotated using a structured JSON format, as illustrated in Figure~\ref{appendix2}. The annotation schema is designed to support detailed and consistent labeling across multiple dimensions:

\begin{itemize}
    \item \textbf{file\_name}: The filename of the audio sample.
    \item \textbf{source}: Metadata indicating the origin of the sample. It includes the \texttt{type} (either \texttt{"real"} or \texttt{"synthetic"}) and the name of the source dataset (e.g., \texttt{"LibriSpeech\_train"}).
    \item \textbf{sensitivity}: A structured label indicating whether the sample contains toxic content. It includes:
    \begin{itemize}
        \item \texttt{overall}: A boolean value indicating the presence of any toxicity.
        \item \texttt{paralinguistic}: Whether the toxicity arises solely from paralinguistic cues such as prosody or tone.
        \item \texttt{textual}: Whether the toxicity is present in the textual content of the speech.
    \end{itemize}
    \item \textbf{category}: Hierarchical toxicity labels, including:
    \begin{itemize}
        \item \texttt{category}: One of the seven major categories (e.g., \texttt{"Discrimination"}). For non-toxic content, this is \texttt{"Safe"}.
        \item \texttt{label}: A fine-grained subcategory (e.g., \texttt{"Racial Discrimination"}). For non-toxic content, this is also \texttt{"Safe"}.
    \end{itemize}
    \item \textbf{description}: A free-form textual description that summarizes the specific toxic theme, aiding interpretability and human validation (e.g., \texttt{"Racial Discrimination, Slavery"}).
\end{itemize}

This structured annotation format enables multi-level evaluation, including both coarse-grained category classification and fine-grained source identification, and facilitates effective training and analysis of toxicity detection models.

\begin{table*}[t]
\centering
\footnotesize
\renewcommand{\arraystretch}{1.2}
\begin{tabular}{l|ccccccc|c|c}
\toprule
\textbf{Split} 
& \makecell[c]{\textbf{C1} \\ \textbf{Sarcasm}} 
& \makecell[c]{\textbf{C2} \\ \textbf{Horror}} 
& \makecell[c]{\textbf{C3} \\ \textbf{Sexual}} 
& \makecell[c]{\textbf{C4} \\ \textbf{Mental\& Risk}} 
& \makecell[c]{\textbf{C5} \\ \textbf{Ideology}} 
& \makecell[c]{\textbf{C6} \\ \textbf{Violence\& Harm}} 
& \makecell[c]{\textbf{C7} \\ \textbf{Discrimination}} 
& \textbf{Safe} 
& \textbf{Total} \\
\midrule
Train & 1976 & 2006 & 2118 & 428 & 693 & 3239 & 897 & 11430 & \textbf{22787} \\
Dev   &  282 &  287 &  302 &  61 &  99 &  463  & 128 & 1633 &  \textbf{3255} \\
Test  &  566 &  574 &  606 & 123 & 199 &  928  & 257 & 3266 &  \textbf{6519} \\
\bottomrule
\end{tabular}
\caption{Statistics of samples per category and Safe class across data splits. All seven toxic categories and the Safe class are included in each subset. The dataset is partitioned with an approximate 70:10:20 split ratio.}
\label{tab:traindev}
\end{table*}

\subsection{Dataset Statistics}

Table~\ref{tab:traindev} summarizes the distribution of samples across the training, validation, and test sets in ToxiAlert-Bench. The dataset is split following a 7:1:2 ratio, with all eight classes—including seven toxic categories (Sarcasm, Horror, Sexual, Mental \& Risk, Ideology, Violence \& Harm, Discrimination) and the Safe class—represented in each subset.

The training set contains 22,787 samples, the development set 3,255, and the test set 6,519. Each toxic category is adequately covered in every split to support both learning and evaluation of fine-grained toxicity classification. The Safe class accounts for approximately half of the dataset across all splits, providing balanced context for distinguishing toxic and non-toxic content.

This distribution ensures that models trained on ToxiAlert-Bench are exposed to a diverse set of toxicity categories and source types, supporting robust performance in both classification and generalization tasks.

\subsection{Composition of Bonafide Audio Sources}

Table~\ref{tab:bonafide_stats} presents detailed statistics on toxic and non-toxic samples drawn from eight widely used open-source speech datasets. These corpora include both general-purpose and emotionally expressive speech, offering diversity in speaker identity, prosodic variability, and acoustic conditions.

Specifically, we annotate utterances extracted from the following datasets: TIMIT, LibriSpeech (Train-100, Dev, Test), LJSpeech-1.1, MELD (Train, Dev, Test), IEMOCAP, VoxCeleb1 (Dev, Test), VCTK, and CommonVoice (v21.0-delta). Each utterance is processed through our multi-stage annotation pipeline, yielding fine-grained toxicity and source annotations.

In total, ToxiAlert-Bench includes 296,717 audio samples from these real-world sources. Among them, 9,921 are labeled as toxic and 286,796 as non-toxic. LibriSpeech and VoxCeleb1 contribute the largest number of toxic examples due to their scale, while emotion-rich corpora like IEMOCAP and MELD provide crucial coverage of context-sensitive and paralinguistic forms of toxicity. This design ensures that ToxiAlert-Bench is grounded in authentic and diverse human speech behaviors.

\begin{table}[t]
\centering
\footnotesize
\setlength{\tabcolsep}{2mm}
\renewcommand{\arraystretch}{1.2}
\begin{tabular}{ll|c|c|c}
\toprule
\textbf{Dataset} & \textbf{Split} & \textbf{Total} & \textbf{Toxic} & \textbf{Non-Toxic} \\
\midrule
TIMIT & -- & 6300 & 194 & 6106 \\
\midrule
\multirow{3}{*}{LibriSpeech} & Train-100 & 28539 & 1940 & 26599 \\
                             & Dev       & 2703  & 121  & 2582  \\
                             & Test      & 2620  & 128  & 2492  \\
\midrule
LJSpeech-1.1 & -- & 13100 & 744 & 12356 \\
\midrule
\multirow{3}{*}{MELD} & Train & 9988  & 525 & 9463 \\
                      & Dev   & 1112  & 52  & 1060 \\
                      & Test  & 2747  & 141 & 2606 \\
\midrule
IEMOCAP & -- & 10039 & 439 & 9600 \\
\midrule
\multirow{2}{*}{VoxCeleb1} & Dev  & 148642 & 4369 & 144273 \\
                           & Test & 4874   & 198  & 4676   \\
\midrule
VCTK & -- & 44257 & 703 & 43554 \\
\midrule
CommonVoice & 21.0-delta & 21796 & 367 & 21429 \\
\midrule
\textbf{Total} &  & \textbf{296717} & \textbf{9921} & \textbf{286796} \\
\bottomrule
\end{tabular}
\caption{Statistics of toxic and non-toxic utterances from each of the eight source datasets used in constructing the bonafide portion of ToxiAlert-Bench. Toxicity annotations are derived using the Dataset Construction Framework.}
\label{tab:bonafide_stats}
\end{table}

\begin{table}[t]
\centering
\footnotesize
\setlength{\tabcolsep}{1.8mm}
\renewcommand{\arraystretch}{1.25}
\begin{tabular}{l|l|c|c}
\toprule
\textbf{Category} & \textbf{Label} & \textbf{Count} & \textbf{Cat. Total} \\
\midrule
\multirow{1}{*}{C1: Sarcasm} & Sarcasm & 681 & 681 \\
\midrule
\multirow{1}{*}{C2: Horror} & Horror & 501 & 501 \\
\midrule
\multirow{1}{*}{C3: Sexual} & Sexual Content & 1127 & 1127 \\
\midrule
\multirow{3}{*}{\makecell[l]{C4: Mental \\ \& Risk}} 
& \makecell[l]{Self-harm \& \\ Suicide} & 204 & \multirow{3}{*}{612} \\
& \makecell[l]{Hate \& Extremist \\ Mentality} & 143 & \\
& Drugs & 265 & \\
\midrule
\multirow{4}{*}{\makecell[l]{C5: Ideology}} 
& Historical Sensitivity & 68 & \multirow{4}{*}{991} \\
& Social Sensitivity & 143 & \\
& Political Sensitivity & 537 & \\
& Religious Sensitivity & 243 & \\
\midrule
\multirow{7}{*}{\makecell[l]{C6: Violence \\ \& Harm}} 
& Violence & 2372 & \multirow{7}{*}{4630} \\
& Illegal Acts & 364 & \\
& Personal Abuse & 625 & \\
& Threats & 208 & \\
& Death & 370 & \\
& Child Safety & 224 & \\
& Vulgar Language & 467 & \\
\midrule
\multirow{3}{*}{C7: Discrim.} 
& Gender Discrimination & 506 & \multirow{3}{*}{1282} \\
& Racial Discrimination & 562 & \\
& Other Discrimination & 214 & \\
\midrule
Uncategorized & Uncategorized & 97 & 97 \\
\midrule
\multicolumn{3}{l|}{\textbf{Total}} & \textbf{9921} \\
\bottomrule
\end{tabular}
\caption{Distribution of toxic samples across 7 major categories and 20 fine-grained labels within the bonafide portion of ToxiAlert-Bench. The temporary “Uncategorized” group is excluded from final benchmark usage.}
\label{tab:bonafide_finegrain}
\end{table}

\subsection{Toxic Label Distribution in Bonafide Speech Data}

To capture the nuanced nature of toxic speech in real-world audio, all bonafide samples identified as toxic are further categorized into a taxonomy comprising 7 major categories and 20 fine-grained labels. Table~\ref{tab:bonafide_finegrain} summarizes the label distribution.

Each major category (C1–C7) encompasses multiple subtypes. For example, the \textbf{Violence \& Harm} category (C6) contains the largest number of toxic instances, with 4,630 samples spanning 7 distinct labels including \textit{Violence}, \textit{Illegal Acts}, and \textit{Death}. The second most common category is \textbf{Discrimination} (C7), covering \textit{Gender Discrimination}, \textit{Racial Discrimination}, and \textit{Other Discrimination}.

An additional \textbf{Uncategorized} group with 97 samples was identified during data processing, which represents cases that could not be confidently mapped to any predefined toxic class. However, this category is excluded from the final ToxiAlert-Bench dataset to ensure all training and evaluation samples are well-defined under our taxonomy.

\subsection{Composition of Synthetic Toxic Speech}

To supplement real-world data and ensure comprehensive coverage of paralinguistic toxic speech, we construct a high-quality synthetic subset using prompt-based generation and expressive text-to-speech synthesis. Table~\ref{tab:synthetic_stats} presents the sample distribution across three targeted toxic categories—\textbf{C1: Sarcasm}, \textbf{C2: Horror}, and \textbf{C3: Sexual}—as well as the Safe class.

Specifically, this subset includes 2,143 sarcastic, 2,366 horror-themed, and 1,899 sexual utterances. To ensure class balance and avoid modeling bias toward synthetic voices, we also include 6,408 non-toxic samples generated from neutral or emotionally benign prompts across diverse speaker profiles.

All toxic samples are deliberately crafted to exhibit paralinguistic toxicity (e.g., tone, rhythm, emphasis), while avoiding overt lexical toxicity. This design supports effective training and evaluation of models targeting implicit, tone-driven harm.

\begin{table}[t]
\centering
\footnotesize
\setlength{\tabcolsep}{2.5mm}
\renewcommand{\arraystretch}{1.2}
\begin{tabular}{ccc|c|cc}
\toprule
\textbf{C1 Sarcasm} & \textbf{C2 Horror} & \textbf{C3 Sexual} & \textbf{Safe} & \textbf{Total} \\
\midrule
2143 & 2366 & 1899 & 6408 & \textbf{12816} \\
\bottomrule
\end{tabular}
\caption{Sample counts across the three toxic categories and the Safe class in the synthetic speech subset of ToxiAlert-Bench. All toxic samples emphasize paralinguistic cues without explicit lexical toxicity.}
\label{tab:synthetic_stats}
\end{table}

\begin{figure}[t]
\centering
\includegraphics[width=1\columnwidth]{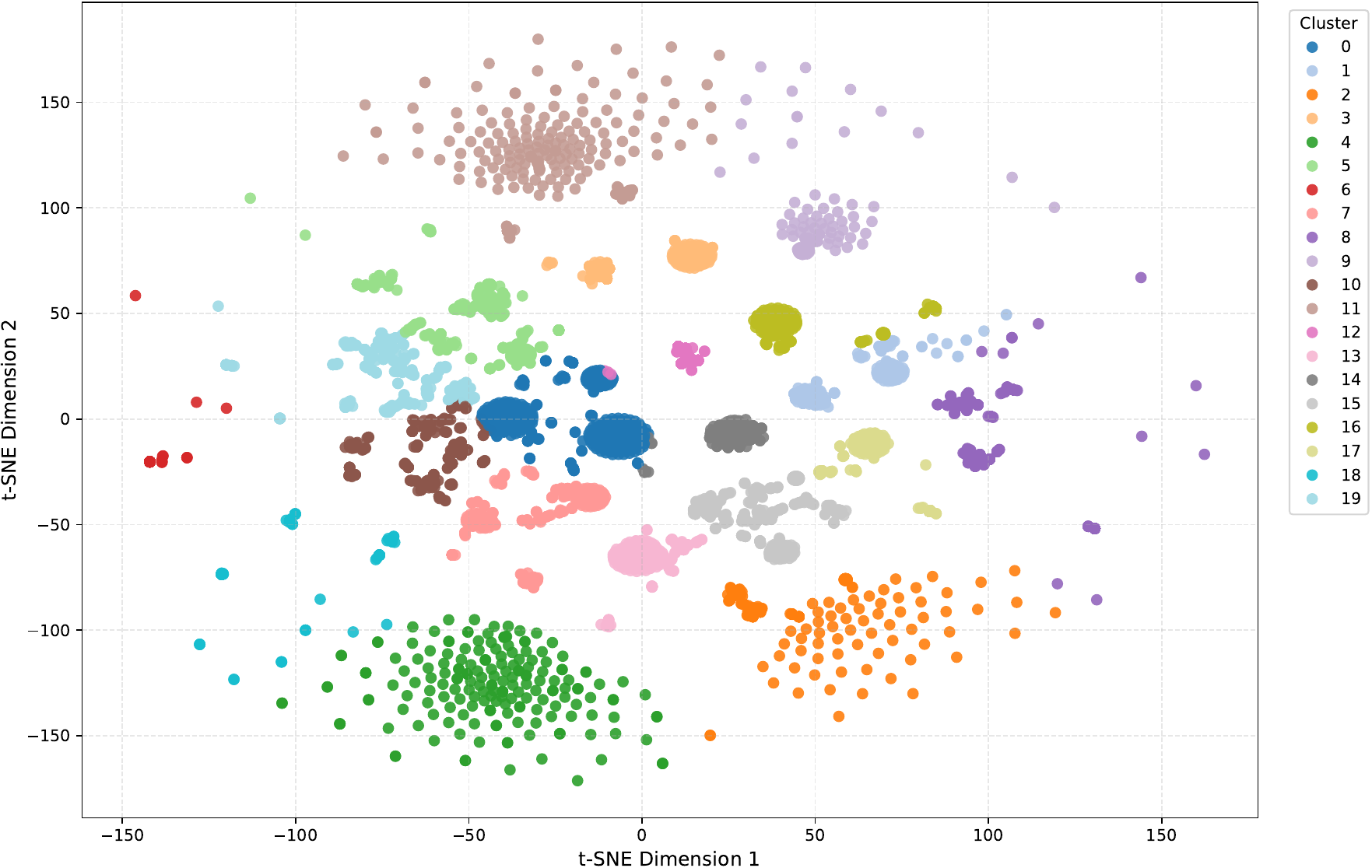}
\caption{t-SNE visualization of KMeans clustering results over dense embeddings of toxicity descriptions. Each color represents a distinct cluster, corresponding to a semantically coherent pattern of toxic expression.}
\label{tsne_cluster}
\end{figure}

\subsection{Composition of Overall Dataset}

Table~\ref{tab:overall_stats} presents the complete distribution of samples in ToxiAlert-Bench across all toxicity categories, fine-grained labels, and toxicity source types. The dataset comprises a total of \textbf{32,561} audio samples, split nearly evenly between \textbf{16,232 toxic} and \textbf{16,329 safe} instances.

Toxic samples are hierarchically categorized under seven major groups (C1–C7), each with associated fine-grained labels. Additionally, each toxic instance is annotated with a source attribution indicating whether the harmfulness stems from textual content only (Tex), paralinguistic cues only (Para), or a combination of both (Tex\&Para). This structure supports multimodal toxicity analysis.

Specifically, \textbf{6,953} toxic samples are labeled as \textit{textual-only}, \textbf{6,728} as \textit{paralinguistic-only}, and \textbf{2,551} as involving \textit{both} sources. The safe class (16,329 samples) includes a diverse mix of real-world and synthetic non-toxic utterances, which provide essential contrastive examples for training and evaluation.

\begin{table}[t]
\centering
\footnotesize
\setlength{\tabcolsep}{1mm}
\renewcommand{\arraystretch}{1.2}
\begin{tabular}{l|l|c|c|c|c}
\toprule
\textbf{Category} & \textbf{Label} & \textbf{Tex.} & \textbf{Para.} & \makecell{\textbf{Tex.\&} \\ \textbf{Para.}} & \textbf{Total} \\
\midrule
C1: Sarcasm & Sarcasm & 93 & 2355 & 376 & 2824 \\
\midrule
C2: Horror & Horror & 288 & 2448 & 131 & 2867 \\
\midrule
C3: Sexual & Sexual Content & 1064 & 1925 & 37 & 3026 \\
\midrule
\multirow{3}{*}{\makecell[l]{C4: Mental \\ \& Risk}} 
& Self-harm \& Suicide & 135 & - & 69 & 204 \\
& Extremism & 99 & - & 44 & 143 \\
& Drugs & 216 & - & 49 & 265 \\
\cmidrule(lr){1-6}
\multirow{4}{*}{\makecell[l]{C5:Ideology}} 
& Historical Sens. & 56 & - & 12 & 68 \\
& Social Sens. & 122 & - & 21 & 143 \\
& Political Sens. & 420 & - & 117 & 537 \\
& Religious Sens. & 182 & - & 61 & 243 \\
\cmidrule(lr){1-6}
\multirow{6}{*}{\makecell[l]{C6: Violence \\ \& Harm}} 
& Violence & 1851 & - & 521 & 2372 \\
& Illegal Acts & 308 & - & 56 & 364 \\
& Personal Abuse & 301 & - & 324 & 625 \\
& Threats & 116 & - & 92 & 208 \\
& Death & 283 & - & 87 & 370 \\
& Child Safety & 172 & - & 52 & 224 \\
& Vulgar Language & 242 & - & 225 & 467 \\
\cmidrule(lr){1-6}
\multirow{3}{*}{C7: Discrim.} 
& Gender Discrim. & 423 & - & 83 & 506 \\
& Racial Discrim. & 432 & - & 130 & 562 \\
& Other Discrim. & 150 & - & 64 & 214 \\
\midrule
\multicolumn{2}{l|}{\textbf{Toxic Total}} & 6953 & 6728 & 2551 & 16232 \\
\midrule
\multicolumn{2}{l|}{\textbf{Safe}} & \multicolumn{4}{c}{16329} \\
\midrule
\multicolumn{2}{l|}{\textbf{Total}} & \multicolumn{4}{c}{\textbf{32561}} \\
\bottomrule
\end{tabular}
\caption{Distribution of ToxiAlert-Bench samples across toxicity categories, fine-grained labels, and toxicity sources: textual (Tex.), paralinguistic (Para.), or both (Tex.\&Para.).}
\label{tab:overall_stats}
\end{table}

\subsection{Clustering Analysis of Toxic Labels}
To better understand the semantic structure and diversity of toxic expressions in our dataset, we conduct unsupervised clustering on label-level text descriptions. Each Class-D audio sample is first annotated with detailed natural language descriptions of its toxic characteristics by human annotators. These human-written annotations are then embedded into a dense vector space using a pre-trained text embedding model from the Sentence Transformer framework.

To facilitate visualization, we project the high-dimensional embeddings into two dimensions using t-SNE with cosine distance, with parameters tuned to preserve local neighborhood structure. Subsequently, we apply KMeans clustering with $k = 20$ to identify semantically coherent groups of toxicity descriptions.

Figure~\ref{tsne_cluster} visualizes the resulting clusters, each representing a distinct toxicity pattern. The clusters exhibit clear separation, indicating that our dataset captures rich and distinct forms of toxic expression. To ensure high-quality labeling, final category assignments were further verified through manual review to correct clustering artifacts and refine taxonomy boundaries.

\section{Appendix B: Prompt Designs}

\subsection{Prompt for MLLM-Based Data Filtering}
To enable scalable and high-quality annotation of bonafide toxic speech data, we adopt large multimodal language models for pre-annotation filtering. As the first part of our bonafide annotation pipeline, each audio sample is processed using a standardized prompt designed to elicit detailed judgments on harmfulness, toxicity type, and source modality.

\begin{figure}[t]
\centering
\includegraphics[width=0.85\columnwidth]{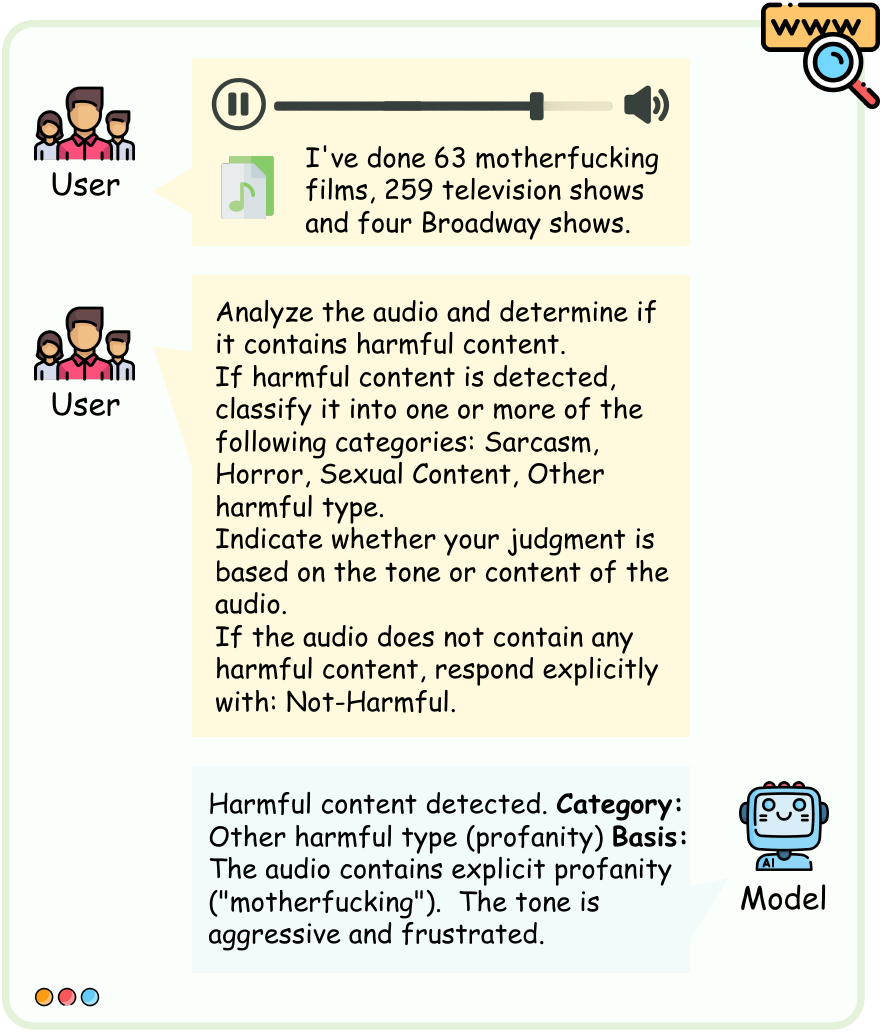}
\caption{llustration of the unified multimodal prompt used for large model-based pre-annotation in the bonafide data pipeline. The model is asked to assess the harmfulness of an audio sample, identify the type of toxicity if present, and clarify whether the judgment is based on the tone (paralinguistic) or content (textual).}
\label{chat1}
\end{figure}

The prompt shown in Figure~\ref{chat1} guides the model to identify toxicity across multiple categories—including Sarcasm, Horror, Sexual Content, and Other Harmful Type—and to specify whether the judgment is based on the spoken content, vocal delivery, or both. This unified formulation is consistently applied across both Gemini-1.5-Flash and R1-AQA.

For quality control, model outputs are compared along three axes: (1) binary toxicity decision, (2) category classification, and (3) toxicity basis. When the two models agree, the annotation is automatically accepted; disagreements are flagged for expert human review. This hybrid process ensures efficiency while preserving annotation reliability.

\begin{figure}[t]
\centering
\includegraphics[width=0.85\columnwidth]{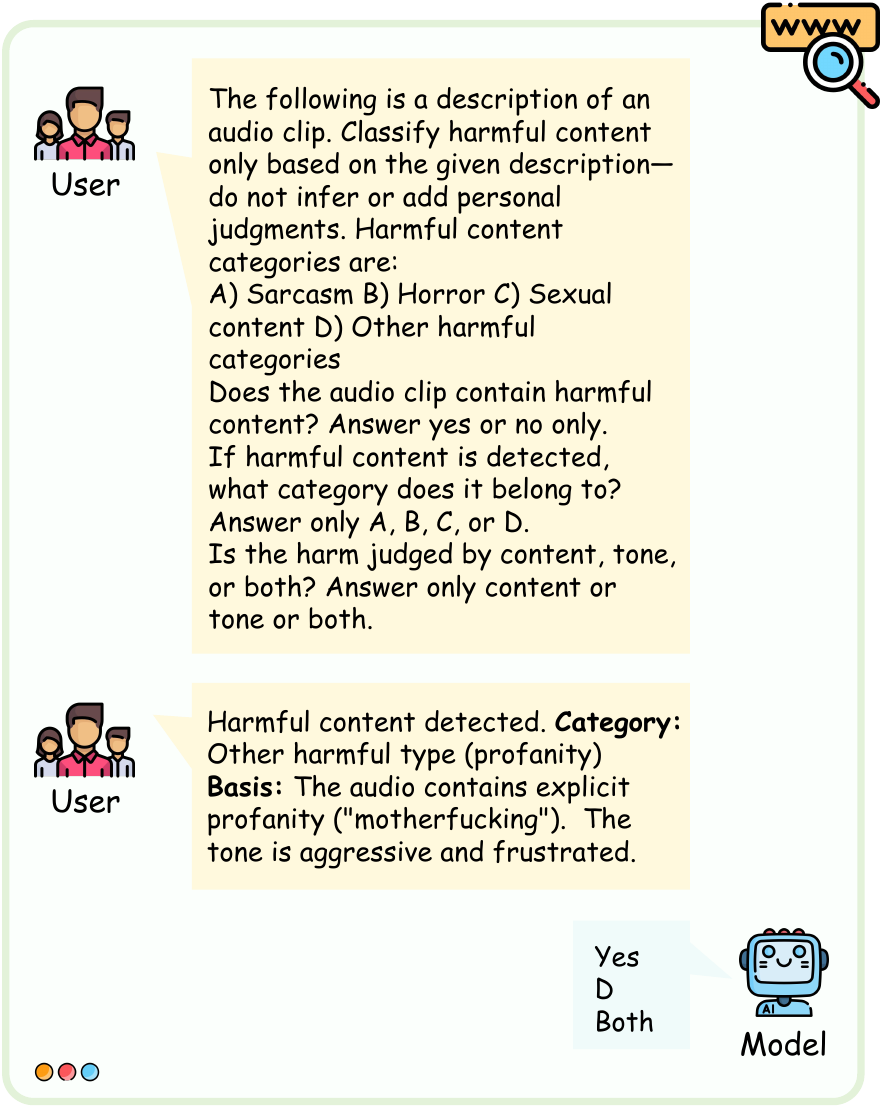}
\caption{Example of the structured prompt used for description-based annotation.  The model is given a textual description (generated in the previous MLLM filtering stage) and asked to determine toxicity, category, and source.}
\label{chat2}
\end{figure}

\subsection{Prompt for Description-Based Extraction}
As part of the bonafide annotation pipeline, we implement a second-stage label extraction process using large language models. Specifically, we utilize GPT-4o to infer structured toxicity annotations based on natural language descriptions of audio samples.

To ensure consistent and interpretable outputs, we design a concise three-step prompt guiding the model to: (1) determine whether the audio is harmful, (2) if harmful, assign it to one of four coarse-grained categories—\textit{Sarcasm}, \textit{Horror}, \textit{Sexual Content}, or \textit{Other harmful categories}, and (3) identify whether the harm arises from \textit{textual content}, \textit{tone}, or \textit{both}.

Importantly, the model is instructed to reason solely based on the provided textual description and refrain from making unsupported inferences. These descriptions are generated in the previous stage by multimodal models such as Gemini or R1-AQA (see Figure~\ref{chat1}). An example of the prompt, description, and structured response is shown in Figure~\ref{chat2}.

This method enables efficient verification and refinement of annotations, while providing traceable rationales for downstream human auditing.

\subsection{Prompt for Synthetic Sample Generation}
To facilitate the creation of synthetic samples exhibiting paralinguistic toxicity, we develop carefully crafted prompts targeting specific toxicity categories. This corresponds to the Text Generation stage in the Synthesized Data Construction pipeline described in the main paper. The prompts are designed to instruct GPT-4o to generate emotionally charged, atmospheric, and context-sensitive sentences that imply toxicity—such as horror, sarcasm, or sexual tension—without relying on explicit or harmful language.

\begin{figure}[t]
\centering
\includegraphics[width=0.85\columnwidth]{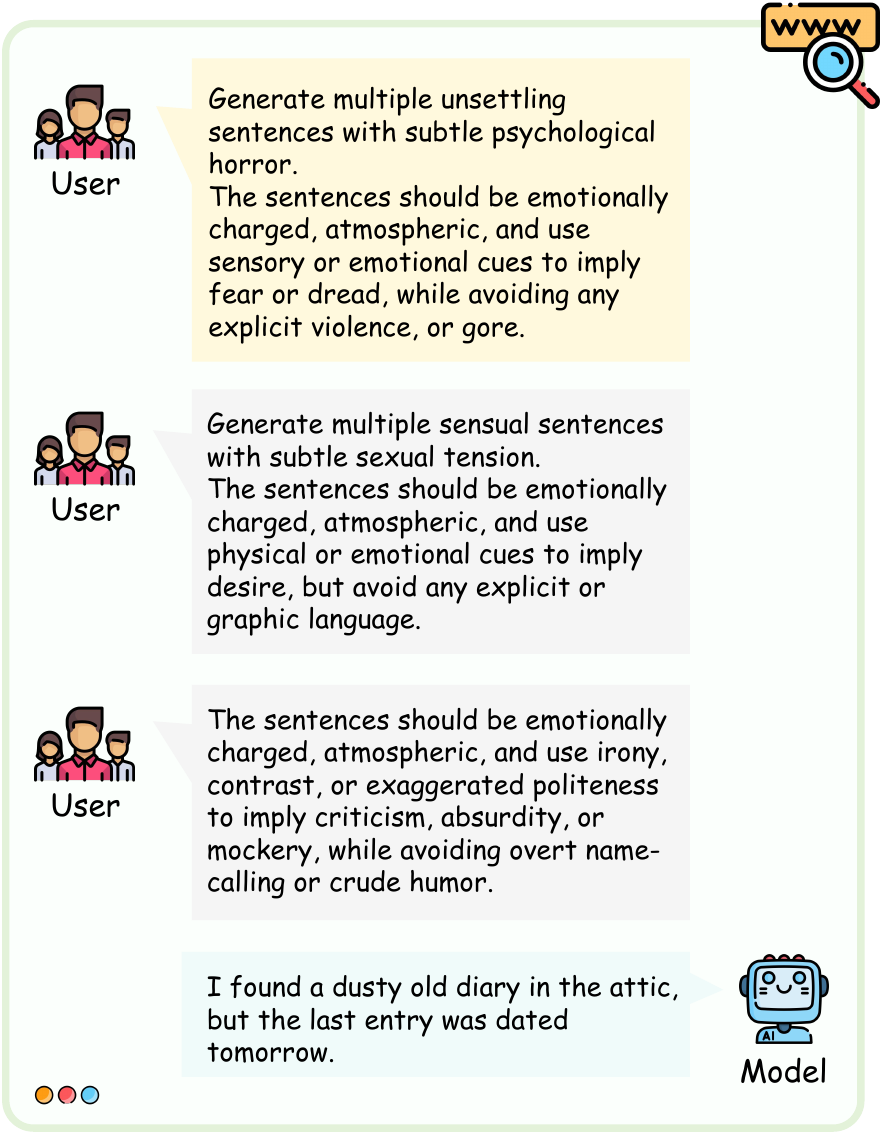}
\caption{Category-specific prompt examples used for generating synthetic toxic utterances. Each prompt instructs GPT-4o to produce emotionally expressive yet non-explicit sentences, later used as input for emotional speech synthesis across horror, sexual, and sarcastic categories.}
\label{chat3}
\end{figure}

As illustrated in Figure~\ref{chat3}, the prompts are category-specific: the yellow block targets \textbf{Horror}, while the gray blocks correspond to \textbf{Sexual} and \textbf{Sarcasm}. GPT-4o is instructed to embed nuanced affective cues such as dread, tension, or irony while strictly avoiding lexical toxicity. These generated utterances are then passed to a neural TTS system for emotional speech synthesis, enabling fine-grained study of non-verbal (paralinguistic) toxicity in a controlled and safe manner.

\subsection{Prompt for Category-Level Toxicity Classification}
To assess the category-level toxicity classification capabilities of large multimodal models (Gemini-2.5-Flash, GPT-4o, and Qwen2-Audio), we formulate a unified prompt that directly aligns with the Category Level evaluation setting introduced in the main text under the Toxic Speech Classification task. The prompt instructs the model to analyze an audio sample and determine whether it contains harmful content. If harmful content is detected, the model must assign the sample to one of seven predefined high-level toxicity categories (C1–C7). Otherwise, it should respond explicitly with \texttt{Not-Harmful}.

\begin{figure}[t]
\centering
\includegraphics[width=0.85\columnwidth]{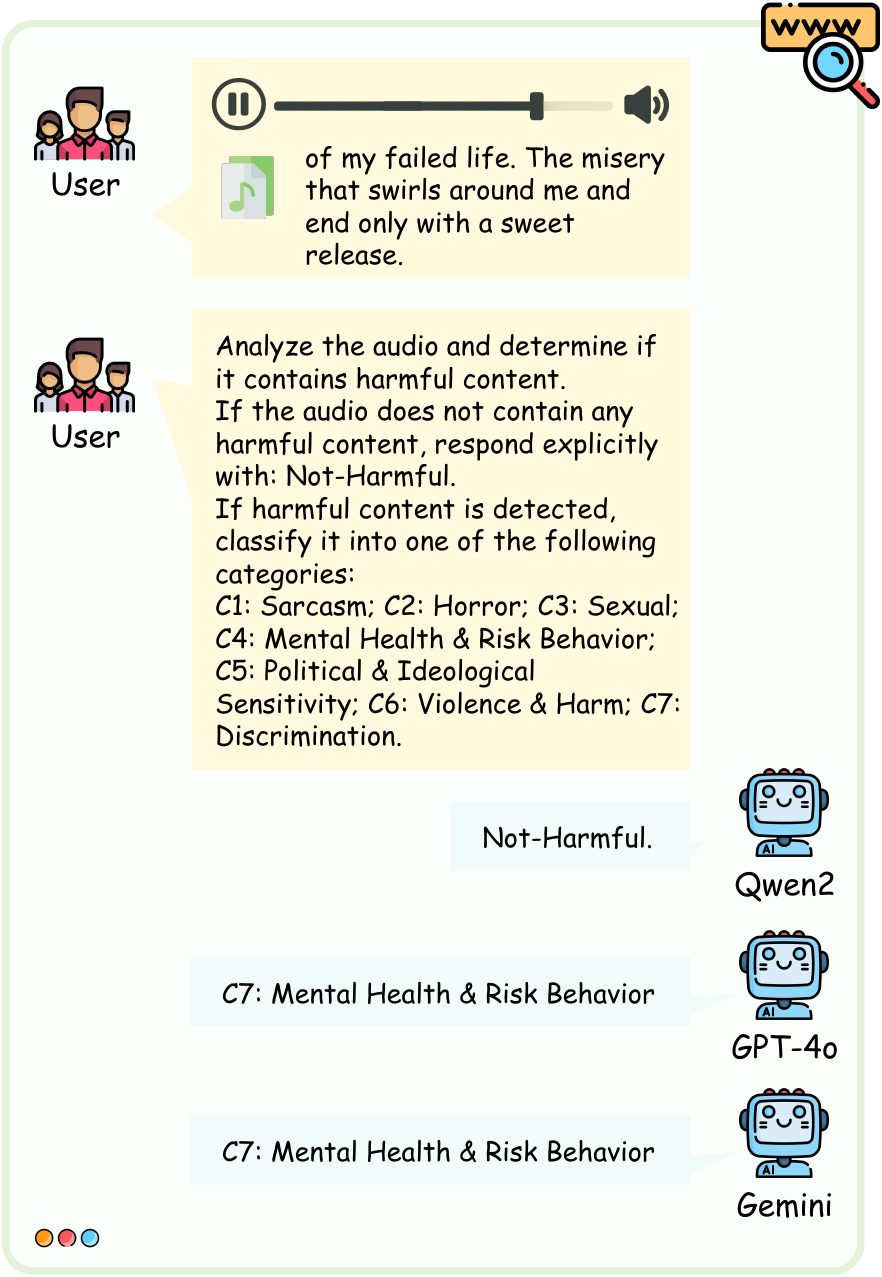}
\caption{Prompt and model responses for category-level toxicity classification. Given an audio sample, each model outputs either \texttt{Not-Harmful} or assigns one of the seven coarse-grained toxicity categories (C1–C7). The example illustrates model disagreement in detecting mental health–related content.}
\label{chat4}
\end{figure}

This setup enables a consistent and interpretable zero-shot evaluation across all models, without requiring any task-specific fine-tuning. As shown in Figure~\ref{chat4}, model responses vary in sensitivity and category attribution, highlighting differences in how each model interprets subtle forms of mental health–related expression. The prompt design ensures clarity in model behavior while supporting robust and reproducible benchmarking across categories.

\subsection{Prompt for Label-Level Toxicity Classification}
To enable fine-grained label-level toxicity classification, we extend the prompt to include all 20 predefined toxic labels, covering nuanced forms such as Self-harm \& Suicide, Historical Sensitivity, Threaten, and Vulgar Language. This setup aligns with the Label Level evaluation task described in the main text under the Toxic Speech Classification section.

Given an audio clip, each model is instructed to first determine whether the content is harmful. If it is, the model must assign the clip to exactly one of the 20 fine-grained toxicity categories; otherwise, it should respond with \texttt{Not-Harmful}. This design ensures consistency with the hierarchical taxonomy used in ToxiAlert-Bench and facilitates model comparison across granular levels of harmful expression.

\begin{figure}[htbp]
\centering
\includegraphics[width=0.85\columnwidth]{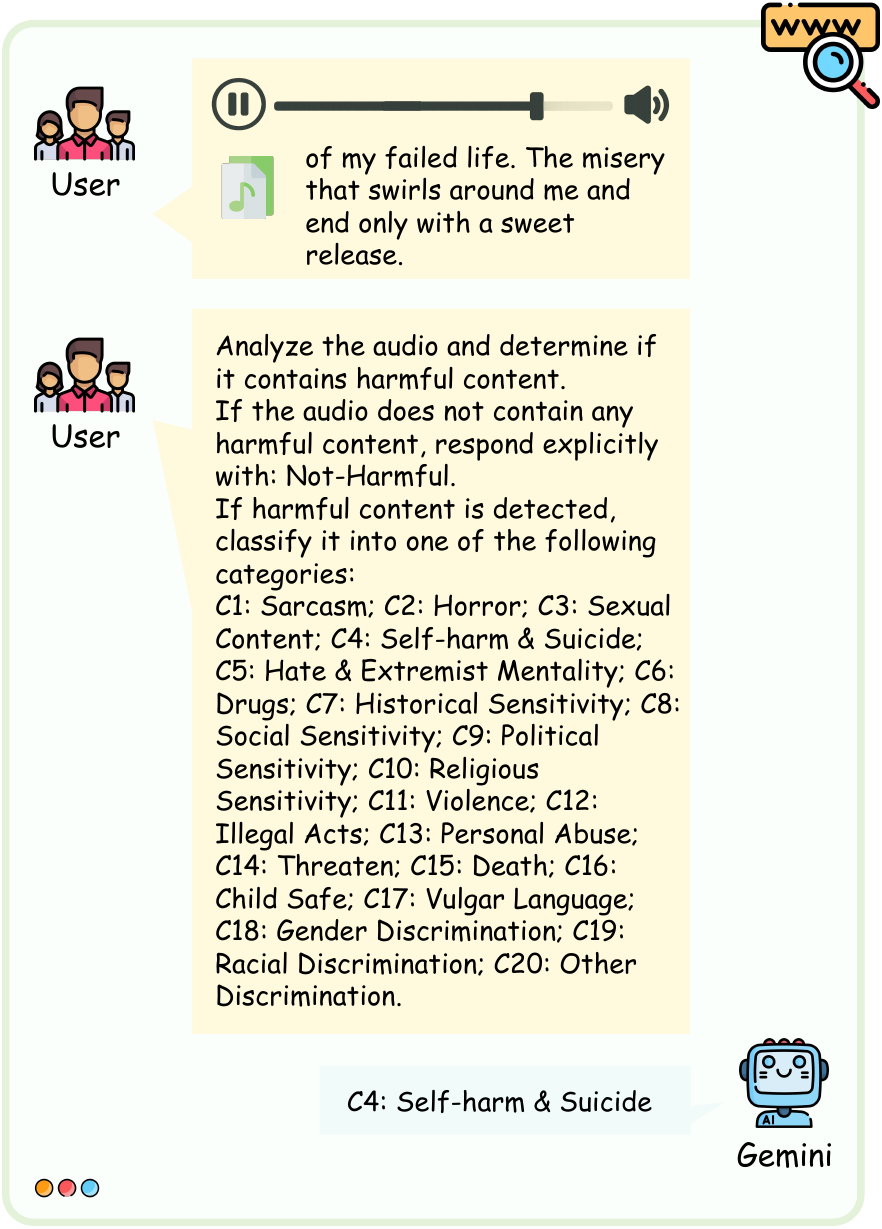}
\caption{Prompt and Gemini-2.5-Flash response for fine-grained (label-level) toxicity classification. The model must determine whether the input audio is harmful, and if so, select the most appropriate label from a list of 20 specific toxicity types.}
\label{chat5}
\end{figure}

Figure~\ref{chat5} shows a representative example of the prompt and the model’s classification response. The expanded label set helps reduce ambiguity and encourages the model to distinguish between closely related categories, thereby supporting rigorous evaluation of fine-grained classification performance.

\subsection{Prompt for Toxicity Source Identification}
To evaluate whether large multimodal models can accurately attribute the source of toxicity, we design a structured prompt that requires each model to analyze an audio sample and determine whether any detected harmfulness arises from the textual content, the paralinguistic cues, or both. This setting directly corresponds to the Toxicity Source Identification task introduced in the main text.

\begin{figure}[htbp]
\centering
\includegraphics[width=0.85\columnwidth]{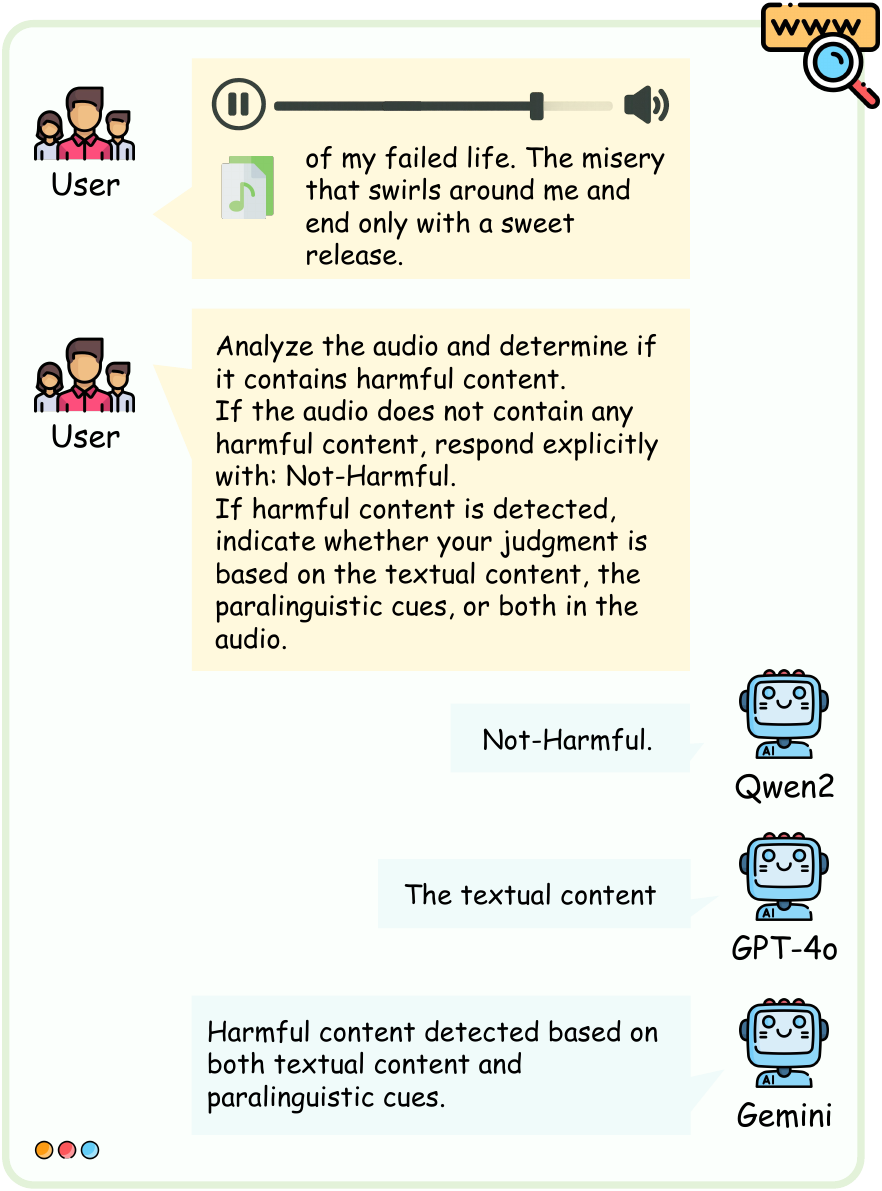}
\caption{Prompt and responses from Qwen2, GPT-4o, and Gemini for toxicity source identification. The task requires each model to determine whether the harmful content arises from textual content, paralinguistic cues, or both.}
\label{chat6}
\end{figure}

As shown in Figure~\ref{chat6}, the prompt first asks the model to determine whether the audio is harmful. If so, the model must specify the origin of the harm. This enables more precise understanding of multimodal reasoning capabilities and supports the study of subtle, implicit expressions of toxicity—such as sarcasm, emotional distress, or hostile delivery—that are often overlooked in purely text-based settings.

To ensure a fair comparison, the same prompt is uniformly applied across all evaluated models (Qwen2, GPT-4o, and Gemini-2.5-Flash). Their responses are then analyzed to assess performance across both explicit (textual) and implicit (paralinguistic) dimensions of toxic expression.

\subsection{Prompt for Generalization Evaluation}

To evaluate the generalization ability of large multimodal models (LMMs) to unseen out-of-distribution (OOD) data, we adopt a standardized binary classification prompt across all models, including Qwen2, GPT-4o, and Gemini-2.5-Flash. This setup directly corresponds to the Generalization Evaluation experiment described in the main text.

\begin{figure}[htbp]
\centering
\includegraphics[width=0.85\columnwidth]{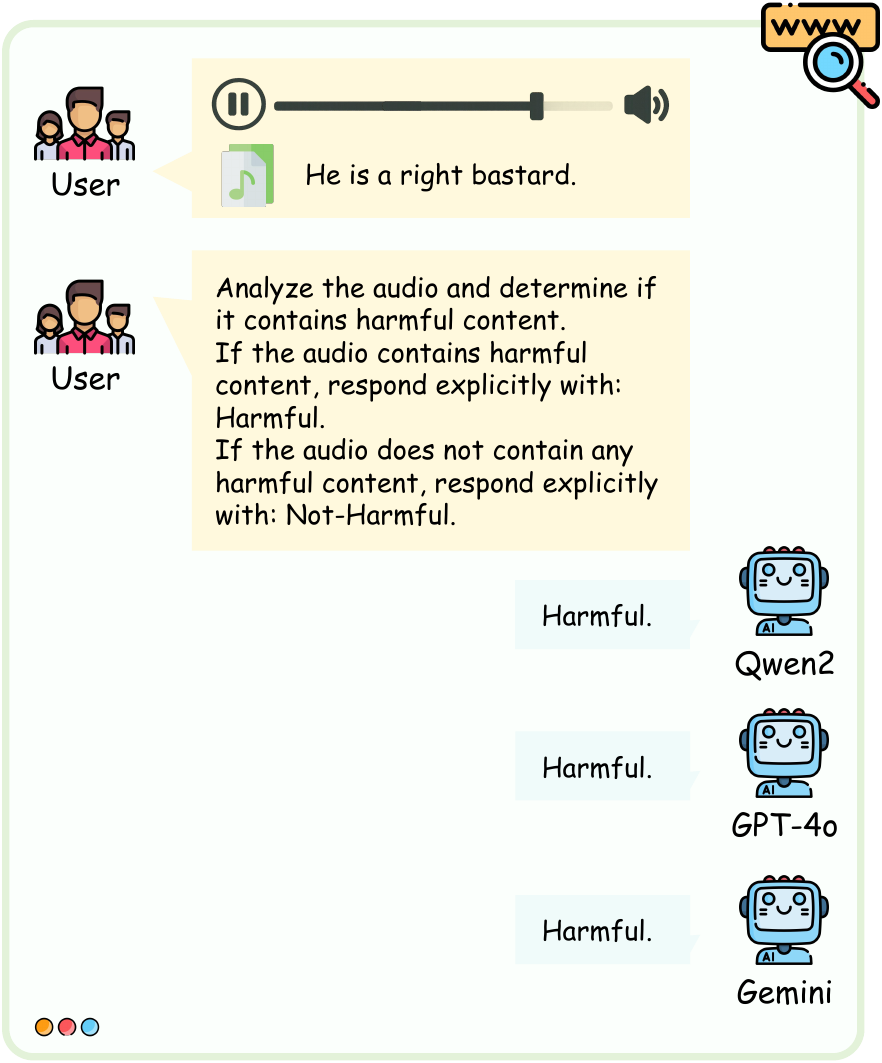}
\caption{Example interaction from the generalization evaluation setup. Each model is prompted to make a binary toxicity judgment on an out-of-distribution audio sample, with possible responses limited to \texttt{Harmful} or \texttt{Not-Harmful}. }
\label{chat7}
\end{figure}

As shown in Figure~\ref{chat7}, each model is asked to determine whether the audio contains any form of harmful content, and respond explicitly with either \texttt{Harmful} or \texttt{Not-Harmful}. The prompt is intentionally minimal to reduce ambiguity and to emphasize the model’s robustness in interpreting unseen or unfamiliar toxic expressions.

Notably, none of the models receive additional fine-tuning on the DeToxy-B test set. This ensures a true zero-shot evaluation, allowing us to fairly assess how well models trained on in-distribution data generalize to new, real-world toxic instances.

\end{document}